\documentclass[12pt,english]{article}
\usepackage{lmodern}
\usepackage[numbers]{natbib}
\usepackage{hyperref}
\usepackage{url}

\usepackage[T1]{fontenc}
\usepackage[latin9]{inputenc}
\usepackage{geometry}
\usepackage{color}
\usepackage{graphicx}
\usepackage{babel}
\usepackage{subfig}

\begin{document}
\hypersetup{
 linkcolor=blue,urlcolor=blue,citecolor=blue}
 
\date{19$^{\mathrm{th}}$ October 2023}
\title{Static Blackhole with Cosmological Influence: Whittaker Solutions}
\author{ Santanu Tripathy and K Rajesh  Nayak  \\
IISER-Kolkata.}
\maketitle
\begin{abstract}
In this article, we investigate the impact of cosmological parameters on black holes using an exact solution to Einstein's equations that satisfies the Whittaker equation of state. We examine a spherically symmetric black hole in the background of a static Einstein Universe with a perfect fluid source with cosmological constant. This solution is characterized by two independent parameters, namely the size of the universe~($R$) and the cosmological constant~($\Lambda$), which represent the cosmological influences. We explore phenomena such as periastron precession and the scattering of massless scalar fields to determine how these cosmological parameters affect the physics around black holes. 
\end{abstract}
\section{Introduction}
Blackhole physics is an intriguing field within both the framework of general 
relativity and alternative theories of gravitation. Extensive investigations have been conducted on classes of solutions, such as Schwarzschild and Kerr black holes~\cite{1916AbhKP1916..189S, schwarzschild1999gravitational,PhysRevLett.11.237}. These solutions exhibit 
characteristics such as a flat spacetime far from the blackhole's event horizon 
and represent vacuum solutions completely devoid of matter. In most studies, 
the cosmological influences on astrophysical blackhole are ignored.   However, 
in reality, we expect the presence of matter surrounding astrophysical 
blackholes or a spacetime that may display cosmological properties as one 
moves away from the event horizon~(EH).

Another crucial property often essential in blackhole physics is the existence of 
time symmetry or a timelike Killing-vector. Though concepts such as 
isolated horizon and dynamical horizon exist, without time symmetry, it 
becomes challenging to define the event horizon at each moment in 
time~\cite{10.1063/1.1664717,Ashtekar_2004}. Since the event horizon is a 
global feature of spacetime, it necessitates complete knowledge of the solution 
at all time slices, especially for light propagation. To investigate 
blackholes in cosmological backgrounds, it may be necessary to relax some of 
the conditions mentioned above. Such models could  provide valuable insights 
into the properties of blackholes that exhibit non-flat asymptotic behaviour and 
time dependent solutions.
 
When it comes to the physics of static or stationary non-asymptotically flat 
blackholes, numerous studies have been conducted on the 
Schwarzschild-de-Sitter~(SdS) solution~\cite{1950SRToh..34..160N, Nariai1999}. The SdS 
spacetime is a straightforward extension of the Schwarzschild solution, 
incorporating a cosmological constant, $\Lambda$, but it remains a vacuum 
solution without any matter field content. Vaidya introduced a solution to 
Einstein's equations with blackholes in the background of the Static Einstein 
Universe(SEU)~\cite{Vaidya1977}. This solution includes a matter field generated by 
a perfect fluid. Another possibility involves a fully evolving cosmological 
background, resulting in a time-dependent solution, such as McVittie's solution 
and its generalisation~\cite{10.1093/mnras/93.5.325,PhysRevD.58.064006,PhysRevD.76.063510,PhysRevD.78.024008}. 
In these cases, only a Cauchy horizon or apparent horizon can be constructed 
at any given time. To determine the location of the event horizon, a complete 
solution is required at all times. We focus solely on time-independent solutions 
in this article.

In this work, we investigate some of the static solutions to Einstein's equations 
that have a blackhole event horizon surrounded by matter, while the spacetime 
is finite or asymptotically non-flat in general. These solutions are inspired by 
the solution given by Vaidya, which consists of a Kerr-like event horizon in the 
Einstein static universe~(Vaidya-Einstein-Kerr or 
VEK)~\cite{Vaidya1977,ramachandra2003kerr}. A special case of the VEK 
solution is obtained by setting the rotational parameter, $a$, to zero, resulting 
in the  static spherically symmetric  Vaidya-Einstein-Schwarzschild(VES) 
solution, which has been extensively investigated~\cite{PhysRevD.63.024020, BSR2002, vishveshwara2000black}.
Although the VES solution is intriguing, it has one drawback: a naked 
singularity at the outer boundary of the spacetime. While this solution can be 
matched with the SEU,  the first derivative cannot be smoothly matched, 
making these discontinuities unphysical~\cite{PhysRevD.63.024020}. One 
interesting aspect of the VES solution is that the density and pressure satisfy 
the condition $\rho+3p=0$, which corresponds to a special case of the 
Whittaker equation of 
state~\cite{stephani_kramer_maccallum_hoenselaers_herlt_2003}.

Though the VES solution is very interesting, it has one drawback: a naked 
singularity  at the outer  boundary of the spacetime. While this solution can be 
matched with the  SEU, though metric coefficient can be matched, the first 
derivate  can not be matched smoothly, this discontinuities make the matching 
unphysical~\cite{PhysRevD.63.024020}. 
One interesting point about  VES solution is that the density and pressure 
satisfy the  condition $\rho+3p=0$, which is a special  case of the Whittaker 
equation of  
state~\cite{stephani_kramer_maccallum_hoenselaers_herlt_2003}.

The VES solution is a subset of the Whittaker solutions. We extend 
this idea further and seek a solution to Einstein's equations that satisfies the 
Whittaker equation, including the cosmological constant~$\Lambda$. While 
Vaidya's solution includes $\Lambda$, it appears merely as an additive 
constant and does not explicitly appear in the metric coefficients. In this paper, 
we extend Vaidya's solution to include the cosmological constant, resulting in a 
more general solution that explicitly depends on $\Lambda$. We utilise the 
condition $\rho+3p=0$, which is a special case of the Whittaker equation of 
state. This introduces a cosmological horizon~(CH) in addition to the regular 
blackhole event horizon. The CH blocks the naked singularity at the outer 
boundary. This CH is similar to the one present in the de-Sitter type of solution, 
although de-Sitter solutions do not have any curvature singularity beyond the 
cosmological horizon. With this solution, we explore the properties of static 
event horizons in the presence of matter and asymptotically non-flat 
spacetime. Now, we have a generalisation of the SdS solution with matter, a 
blackhole horizon, and a cosmological horizon.

In the case of VSD, the cosmological influence on the event horizon is 
determined by a single parameter, $R$, which can be associated with the size 
of the SEU. As the limit $R\rightarrow \infty$, the spacetime solution matches 
with the Schwarzschild solution.
On the other hand, the Schwarzschild-Whittaker (SW) solution, as presented 
here, has two parameters that can contribute to cosmological influence, 
namely, $R$ and $\Lambda$. Consequently, we can now study the effect on 
the horizon with two additional parameters, in addition to the mass. In the 
limiting case of large $R$ and small $\Lambda$, the SW solution matches with 
the SdS spacetime.

Using the solution to Einstein's equations with a blackhole surrounded by 
matter and an asymptotically non-flat background, we first examine the 
behaviour of the event horizon and its existence with various parameters in the 
solutions. Generally, there can be two horizons, as in the case of SdS 
spacetime. However, in extreme cases, the inner and outer horizons merge, 
and the spacetime become unphysical due to the break down of timelike static 
fluid model.

We then investigate the periastron precession of elliptical orbits. The 
phenomenon of periastron precession has served as an important test to 
understand the behaviour of gravity. Since pure Keplerian orbits do not exhibit 
precession, it has provided a robust test for the general theory of relativity 
itself. However, in the current context, we use orbital precession to understand 
the influence of cosmological parameters on the orbits. We compare the orbital 
precession of the SW spacetime with that of Schwarzschild and SdS 
spacetime.

This article is organised as follows: In Section~\ref{sec:SolveEE}, we solve the 
Einstein's equations with a perfect fluid solution and $\Lambda$. We assume 
that the fluid satisfies the equation of state $\rho+3p=0$. We are not aware of 
any earlier report of such a solution to Einstein's equations in the literature.
In Section~\ref{sec:EHand_SO}, we briefly discuss the properties of the EH 
and static observers. The existence of the horizon depends on the values of 
the parameters $R$ and $\Lambda$.
In Section~\ref{sec:geo_prec}, we discuss the details of geodesic motion and 
periastron precession. Our goal here is to compare the results with the 
standard Schwarzschild blackhole rather than the actual measurement of 
periastron precession. Therefore, we use a numerical method to compute the 
precise value of the precession closer to the blackhole.
In Section~\ref{sec:sch-wth}, we study the behaviour of scalar waves. Finally, 
we close the paper with brief concluding remarks. 
\section{Schwarzschild-Whittaker Solution\label{sec:SolveEE}}
In this section, we extend the VES solution with the Whittaker equation of state, ensuring it satisfies the Einstein field equations with the cosmological constant $\Lambda$. We begin with a modified form of the VES metric, given by,
\begin{equation}
ds^{2}=F\left(r\right)dt^{2}-\frac{dr^{2}}{F\left(r\right)}-R^{2}\sin^{2}\left(\frac{r}{R}\right)d\Omega^2\,,\label{eq:ds2}
\end{equation}
where,
$$
d\Omega^2\,=\,\left(d\theta^{2}+\sin^{2}\theta d\phi^{2}\right)\,.
$$
If $F(r)=1$, the metric matches with the SEU, and the parameter $R$ represents the size of the universe. The cosmological influence here is introduced by the spherical part of the metric only. Here, the smaller the value of $R$ is, the larger the cosmological influence is, and as $R\rightarrow\infty$, the cosmological effect vanishes. Furthermore, we assume the form of $F(r)$,
\begin{eqnarray}
F\left(r\right)&=&\left[1-\frac{2m}{R\tan\left(\frac{r}{R}\right)}-f(r)\right]\,. \label{eq:ar}
\end{eqnarray}
 We obtain a suitable function $f(r)$ that satisfies Einstein's equations. By setting $f(r)=0$, we recover the VES solution. The ranges of coordinates are as follows:
 $0\leq\frac{r}{R}\leq\pi$, $0\leq\theta\leq\pi$, $0\leq\phi\leq2\pi$.
The Einstein field equations, including the cosmological term $\Lambda$, are given by:
\begin{equation}
G_{\:b}^{a}\:=\:R_{\:b}^{a}-\frac{1}{2}\delta_{\:b}^{a}R\ =\ \kappa T_{\:b}^{a}+\Lambda\delta_{\:b}^{a},\label{eq:eneq}
\end{equation}
The non-vanishing components of the Einstein tensor for the metric in Eq. (\ref{eq:ds2}) are given by:
\begin{eqnarray}
G_{\:0}^{0} & = & \frac{3}{R^{2}}\left[1-\frac{2m}{R}\cot\left(\frac{r}{R}\right)\right]+\frac{A}{R^{2}}\left\{ R\cot\left(\frac{r}{R}\right)f'+\left[\csc^{2}\left(\frac{r}{R}\right)-3\right]f\right\} \,,\nonumber \\
G_{\:1}^{1} & = & \frac{1}{R^{2}}\left[1-\frac{2m}{R}\cot\left(\frac{r}{R}\right)\right]+\frac{A}{R^{2}}\cot\left(\frac{r}{R}\right)\left[Rf'+\cot\left(\frac{r}{R}\right)f\right]\,,\nonumber \\
G_{\:2}^{2}=G_{\:3}^{3} & = & \frac{1}{R^{2}}\left[1-\frac{2m}{R}\cot\left(\frac{r}{R}\right)\right]+\frac{A}{2R^{2}}\left[R^{2}f''+2R\cot\left(\frac{r}{R}\right)f'-2f\right]\,.\label{eq:Gabs}
\end{eqnarray}
 where $f'=\frac{\partial f}{\partial r}$. The energy-momentum tensor
$T_{ab}$ is taken to be that of a perfect fluid, takes the standard form,
\begin{equation}
T_{\:b}^{a}\ =\ (\rho+p)u^{a}u_{b}\,-\,p\,\delta_{\:b}^{a},\label{eq:tmn}
\end{equation}
 here, $u^{a}$ is the four-velocity of the fluid element, which is timelike. In the case of a static spacetime, the four-velocity is in the direction of the time-like Killing-vector and is explicitly given by: 
\begin{equation}
u^{a}\ =\ \frac{1}{\sqrt{g_{00}}}\delta_{0}^{a}\,.\label{eq:vel}
\end{equation}
 The energy momentum tensor simplifies to, 
\begin{equation}
T_{\:b}^{a}=\mathrm{dia}\left(\rho,\,-p,\,-p,\,-p\right)\,.
\end{equation}
 We substitute the expression for the energy-momentum tensor back into Eq.~(\ref{eq:Gabs}). For a consistent static solution, we expect $G_{1}^{1}-G_{2}^{2}=0$, which can be simplified to:
\begin{equation}
f''=\frac{2 f}{R^2 \sin^{2}\left(\frac{r}{R}\right)}\,.\label{eq:wc}
\end{equation}
 In addition, it can be easily shown that, 
\begin{equation}
\rho+3p=\epsilon\,,\label{eq:wcnd}
\end{equation}
where $\epsilon$ is a constant. 
The Whittaker solutions satisfy the equation of state given by Eq. (\ref{eq:wcnd})~\cite{W1968}. From Eq. (\ref{eq:Gabs}) and Eq. (\ref{eq:wc}), we obtain,
\begin{equation}
\frac{2\Lambda}{\kappa}-\frac{2}{\kappa R^{2}}\left[\frac{f}{\sin^{2}\left(\frac{r}{R}\right)}\,+\frac{R f'}{\cot\left(\frac{r}{R}\right)}\right]=\epsilon\,,
\end{equation}
 and the solution for $f(r)$ can be readily obtained as,
\begin{equation}
f=\frac{\delta}{\tan\left(\frac{r}{R}\right)}-\left(\Lambda-\frac{1}{2}\epsilon\kappa\right)R^{2}\left[1-\frac{r}{R\tan\left(\frac{r}{R}\right)}\right]\,.\label{eq:f}
\end{equation}
Here, $\delta$ is an arbitrary integrating constant. Finally, we can solve for the pressure $p$ and density $\rho$ as,
\begin{eqnarray}
\rho & = & \frac{3g_{00}}{\kappa R^{2}}-\frac{\epsilon}{2}\,,\nonumber \\
p & = & -\frac{g_{00}}{\kappa R^{2}}+\frac{\epsilon}{2}\,.\label{eq:rp1}
\end{eqnarray}
 The density $\rho$ and pressure $p$ vanish on the Killing-horizon only if we choose the condition $\epsilon=0$. For non-zero $\epsilon$, the pressure and density are discontinuous across the horizon and we do not consider in this article.  For spherically symmetric static spacetimes described by the spacetime metric in Eq. (\ref{eq:ds2}), the 
 Killing-horizon is given by the condition $F(r_{h})=0$. Without loss of generality, in Eq. (\ref{eq:f}), we set the parameter $\delta$ to be zero. With these conditions, we obtain,
\begin{equation}
F\left(r\right)=1-\frac{2m}{R\tan\left(\frac{r}{R}\right)}-\Lambda R^{2}\left[1-\frac{r}{R\tan\left(\frac{r}{R}\right)}\right]\,.\label{eq:g00}
\end{equation}
Although $p<0$, the weak energy condition $\rho+p\geq0$ is always satisfied outside the event horizon for this spacetime. This is the form of metric we use for the rest of our analysis.

As mentioned earlier, one of the main problems with this spacetime is that, in the limit as $\frac{r}{R}\rightarrow\pi$, there is a curvature singularity which is not covered by any event horizon. If the outer solution is matched with the static Einstein solution, it leads to a surface with a discontinuity in the first derivatives of the metric. Interestingly, for the spacetime metric in Eq. (\ref{eq:ds2}) with $g_{00}$ given in Eq. (\ref{eq:g00}), it has an outer cosmological horizon, similar to that of the SdS solution, effectively masking the singularity. In the present study, we use this metric to investigate the effect of matter and cosmological influence on the static blackhole.
\begin{figure}[t]
\begin{center}
\subfloat[]{%
	\includegraphics[scale=0.3]{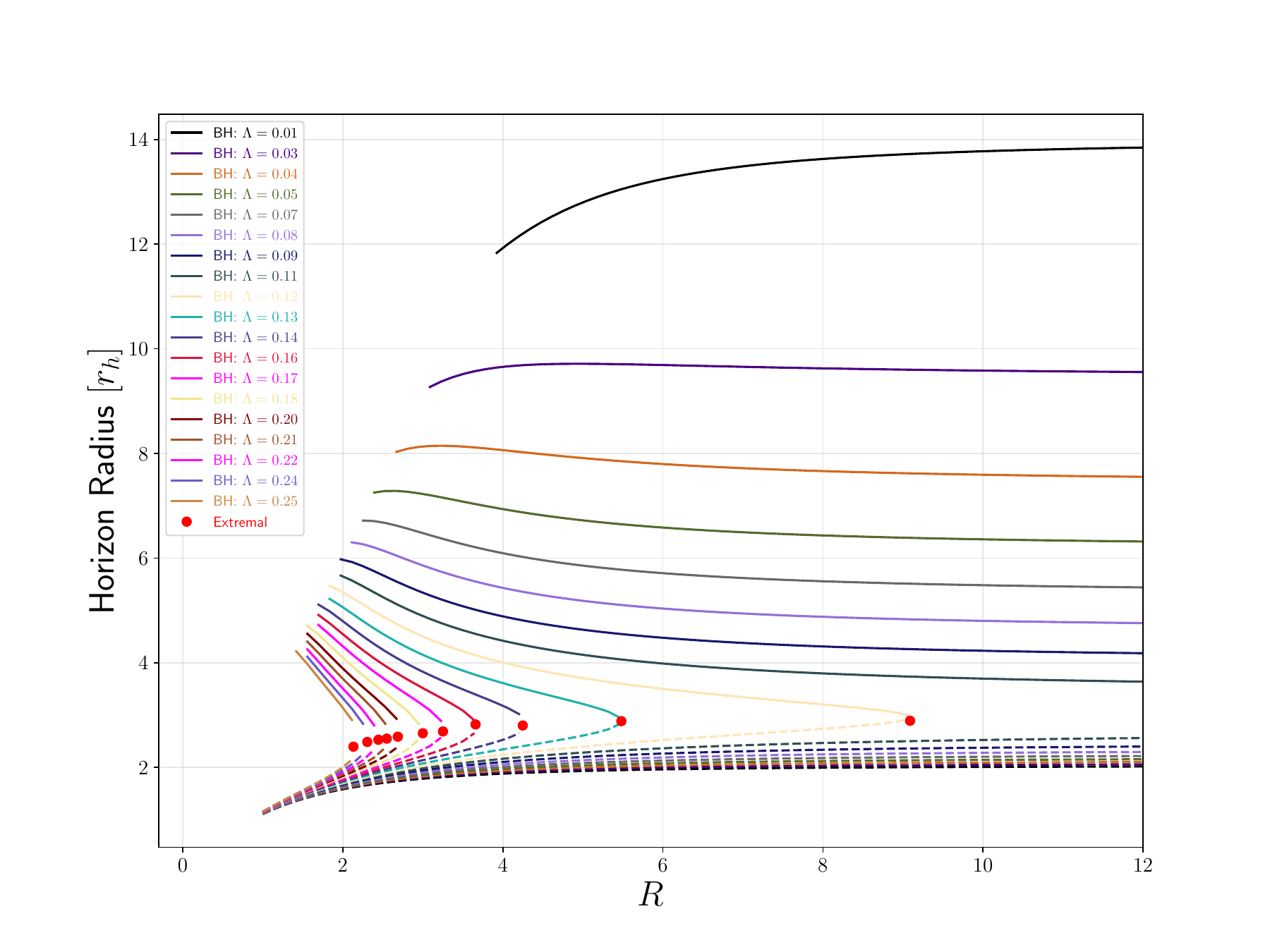}%
        \label{fig:Horizon_funR}%
        }
    \subfloat[]{%
        \includegraphics[scale=0.3]{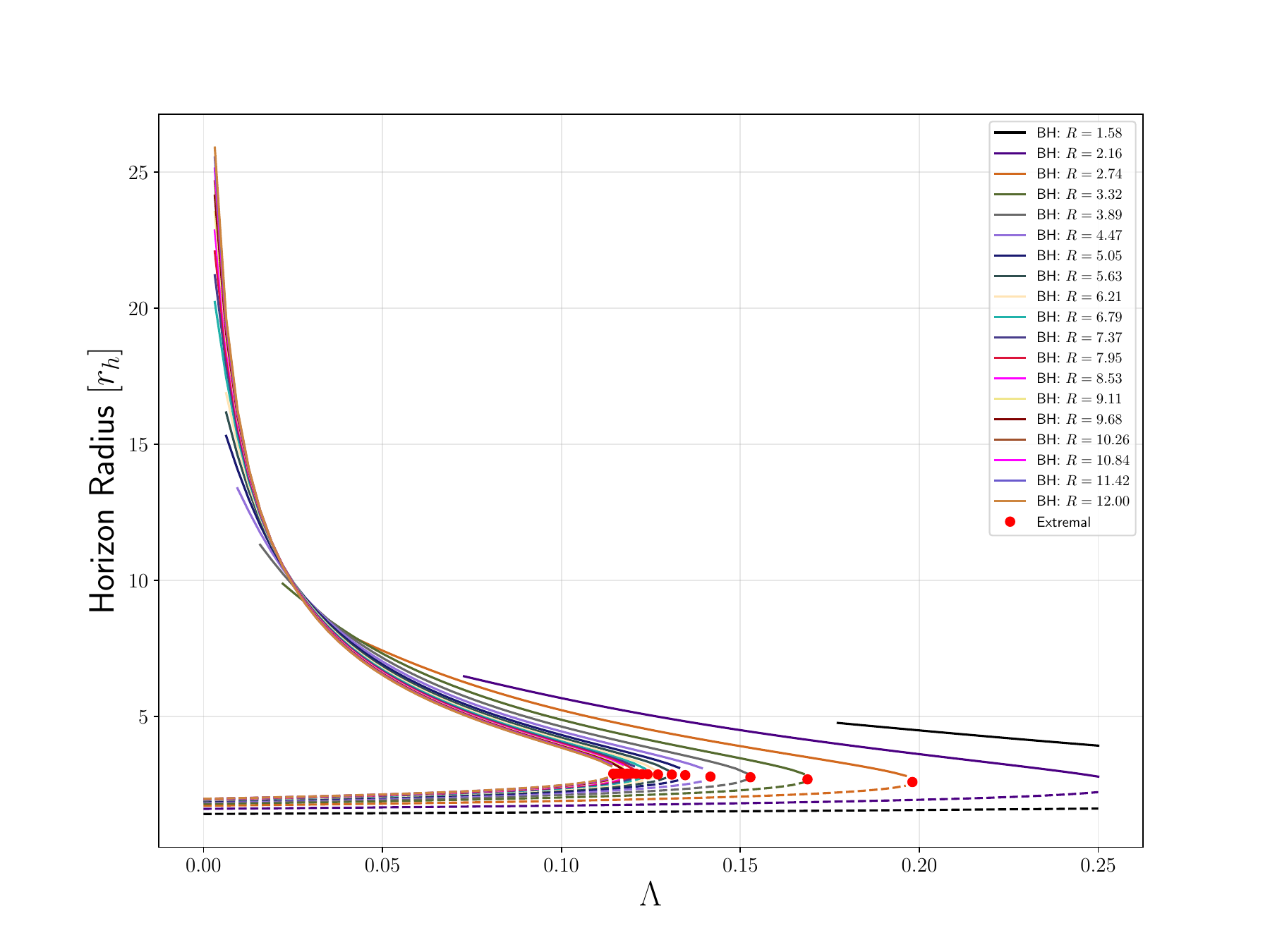}%
        \label{fig:Horizon_funLam}%
        }
\caption{\ \\
(a): Plot of location of the  horizon as a function of $R$ for different values of $\Lambda$. The solid line corresponds to the cosmological horizon (CH), while the dashed line represents the blackhole horizon (BH) for given $\Lambda$ values. The curves for different $\Lambda$ values are shown with different colors. The extremal horizons, where CH and BH merge, are marked with red circles. \\
(b): Plot of location of the  horizon as a function of $\Lambda$ for different values of $R$. The solid line corresponds to the cosmological horizon (CH), while the dashed line represents the blackhole horizon (BH) for given $R$ values. The curves for different $R$ values are shown with different colors. The extremal horizons, where CH and BH merge, are marked with red circles. }
\end{center}
\end{figure}
\subsection{Special cases}
In this subsection, we examine some of the interesting special cases of the spacetime solution obtained in the previous section. We begin with the case when the cosmological constant $\Lambda\rightarrow0$. In this scenario, we obtain the VES spacetime, as given by,
\begin{equation}
ds^{2}=\left[1-\frac{2m}{R\tan\left(\frac{r}{R}\right)}\right]\,dt^{2}-\left[1-\frac{2m}{R\tan\left(\frac{r}{R}\right)}\right]^{-1}\,dr^{2}-R^{2}\sin^{2}\left(\frac{r}{R}\right)d\Omega^2\,.\label{eq:ves}
\end{equation} 
All the additional changes introduced are contributed by the cosmological constant. As the limit $R\rightarrow\infty$, to the lowest order in $r/R$, we obtain the Schwarzschild solution. At the second order in $r/R$, we obtain the SdS spacetime described by the spacetime metric:
\begin{equation}
ds^{2}=\left(1-\frac{2m}{r}-\frac{1}{3}\Lambda r^{2}\right)dt^{2}-\left(1-\frac{2m}{r}-\frac{1}{3}\Lambda r^{2}\right)^{-1}dr^{2}-r^{2}d\Omega^2\,.\label{eq:sch}
\end{equation}
When $m=0$ in eq. (\ref{eq:g00}),
we get another interesting case, where the spacetime metric is given
by, 
\begin{eqnarray}
ds^{2}&=& F_e(r)\,dt^{2}-F_e(r)^{-1}\,dr^{2}
-  R^{2}\sin^{2}\left(\frac{r}{R}\right)d\Omega^2\,,
\end{eqnarray}
Here, $F_e(r)=\left[1-\Lambda R^{2}\left(1-\frac{r}{R\tan\left(\frac{r}{R}\right)}\right)\right]$. This is not de Sitter spacetime because the curvature scalar $K=R^{ijkl},R_{ijkl}$ diverges at both $r\rightarrow0$ and $r\rightarrow R\pi$, similar to the case of VSD spacetime. This interesting special case will be investigated in the future.
\subsection{Another Class of Solution}
Interestingly, the spatial geometry of the SEU can be changed from positive ($+$ve) to negative ($-$ve) by replacing $\sin\left(\frac{r}{R}\right)$ with $\sinh\left(\frac{r}{R}\right)$. The metric given in Eq.~(\ref{eq:ds2}) transforms to, 
\begin{equation}
ds^{2}=H\left(r\right)dt^{2}-\frac{dr^{2}}{H\left(r\right)}-R^{2}\sinh^{2}\left(\frac{r}{R}\right)d\Omega^2\,.\label{eq:dsp2}
\end{equation}
It can be easily shown that in this case, the solution for $H(r)$ takes the form:
\begin{equation}
H\left(r\right)=1-\frac{2m}{R\tanh\left(\frac{r}{R}\right)}-\Lambda R^{2}\left[1-\frac{r}{R\tanh\left(\frac{r}{R}\right)}\right]\,.\label{eq:gp00}
\end{equation}
Proceeding as before, in this case, the expressions for pressure $p$ and density $\rho$ can be written as,
\begin{eqnarray}
\rho & = & -\frac{3g_{00}}{\kappa R^{2}}\,,\nonumber \\
p & = & \frac{g_{00}}{\kappa R^{2}}\,.\label{eq:rp}
\end{eqnarray}
Although the equation of state is still $\rho+3p=0$, this solution does not satisfy the energy condition $\rho > 0$. In an analogous context, this solution could represent a blackhole in an anti-de Sitter background. This solution might be of interest from a cosmological perspective, especially when $m=0$.
\begin{figure}[t]
\begin{center}
\includegraphics[scale=0.28]{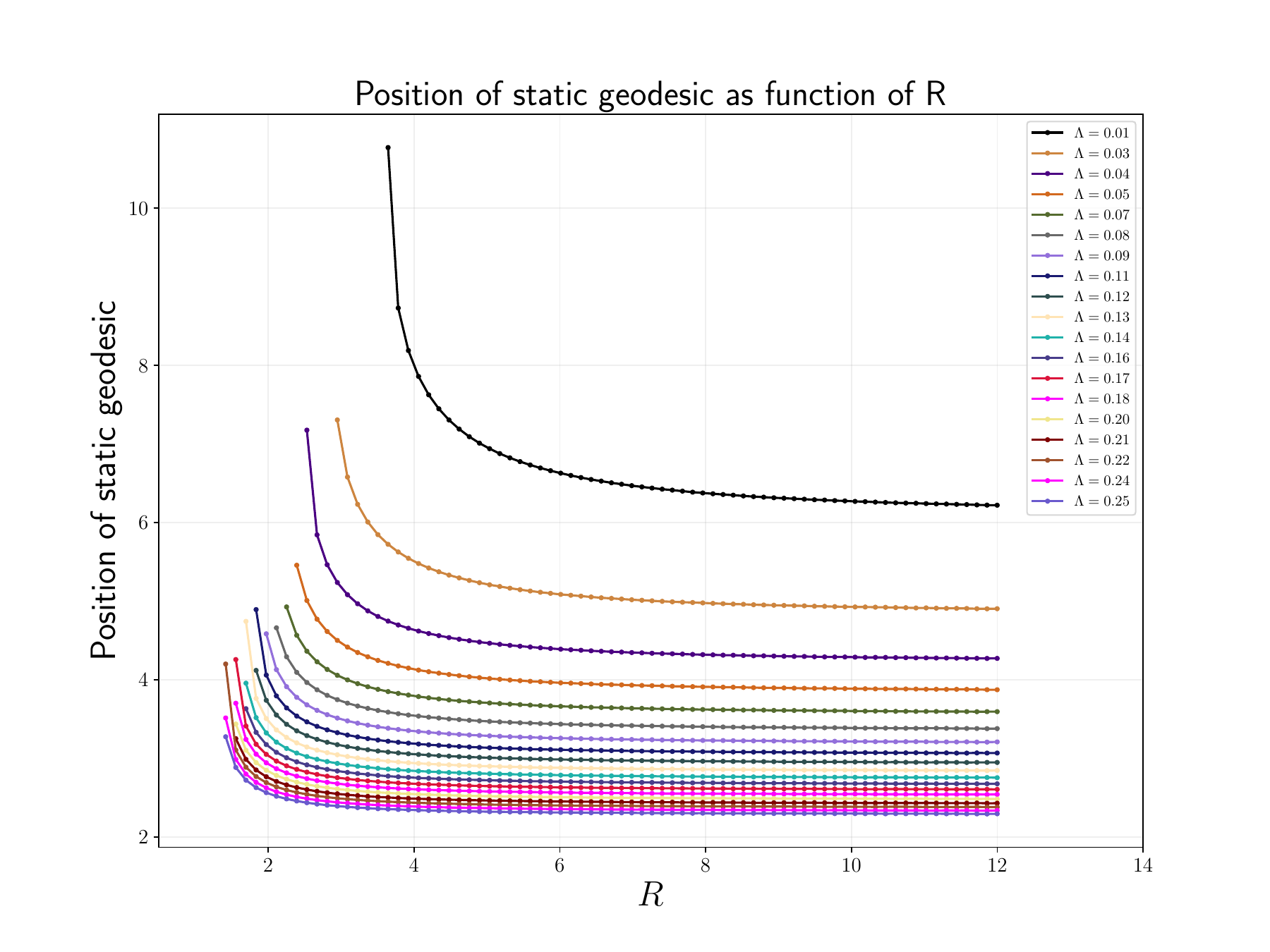}
\includegraphics[scale=0.28]{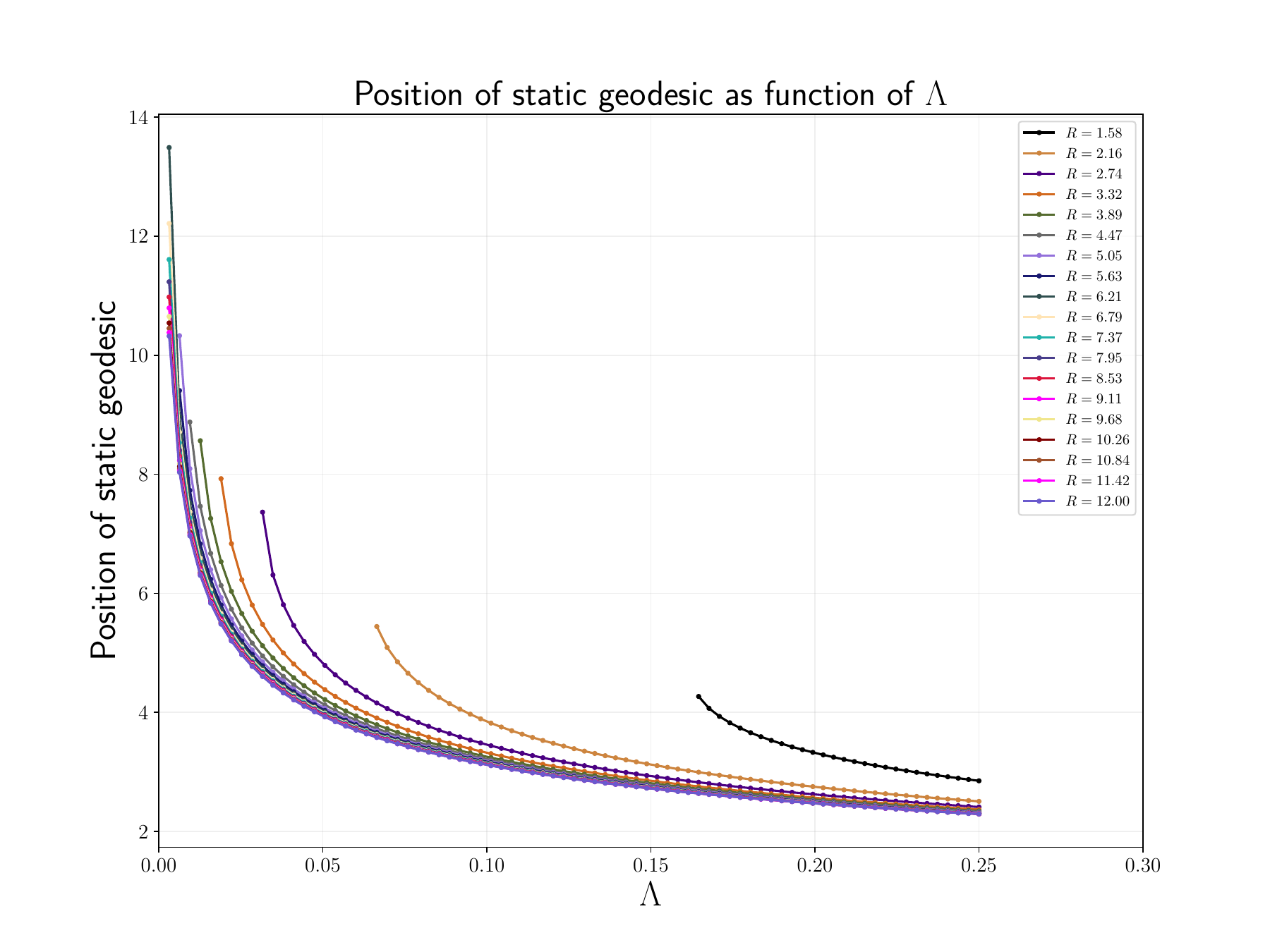}
\caption{Plot of position at which a static observer becomes geodesic.  \label{fig:static_geo}}
\end{center}
\end{figure}
\subsection{Radiating Schwarzschild-Whittaker}
One way to introduce time dependency is by incorporating radiation, similar to a Vaidya-like radiating solution, where the blackhole is embedded in incoming or outgoing radiation~\cite{Vaidya1999}. It turns out that, like the VES solution, the SW solution can be easily extended to incorporate incoming or outgoing radiation~\cite{LKPatel_1979,PhysRevD.31.416}.

We start with the metric given in Eq. (\ref{eq:ds2}), written in the form, 
\begin{equation}
ds^{2}=g_{00}dv^{2}+2\,dr\,dv-R^{2}\sin^{2}\left(\frac{r}{R}\right)\left[d\theta^{2}+\sin^{2}\theta d\phi^{2}\right]\,,
\label{eq:radds}
\end{equation}
 where 
\begin{eqnarray}
g_{00} & \equiv & F\left(v,r\right)=  1-\frac{2\,m\left(v,r\right)}{R\,\tan\left(\frac{r}{R}\right)}-\Lambda R^{2}\left(1-\frac{r}{R\tan\left(\frac{r}{R}\right)}\right)\, , \nonumber \\
v & =& t-r\,.  \ \label{eq:mass_time_relation}
\end{eqnarray}
Here, $m\left(v,r\right)$ is a function of time. In this case, the non-vanishing components of the Einstein tensor are given by,
\begin{eqnarray}
G_{\:0}^{0} & = & \frac{3}{R^{2}}F\left(v,r\right)+\frac{2}{R^{2}}\cot^{2}\left(\frac{r}{R}\right)m'\left(v,r\right) +\Lambda\,, \nonumber \\
G_{\:1}^{0} & = & \frac{2}{R^{2}}\,, \nonumber \\
G_{\:0}^{1} & = & -\frac{2}{R^{2}}\cot^{2}\left(\frac{r}{R}\right) \dot{m}\left(v,r\right)\,, \nonumber \\
G_{\:1}^{1} & = & \frac{1}{R^{2}}F\left(v,r\right)+\frac{2}{R^{2}}\cot^{2}\left(\frac{r}{R}\right)\frac{\partial}{\partial r}m\left(v,r\right)+\Lambda\,,  \nonumber \\
G_{\:2}^{2}=G_{\:3}^{3} & = & \frac{1}{R^{2}}F\left(v,r\right)-\frac{2}{R^{2}}m'\left(v,r\right)+\frac{1}{R}\cot\left(\frac{r}{R}\right)m''\left(v,r\right)\,+\Lambda. \label{eq:eint:radiation}
\end{eqnarray}
 In the above, we have $m'=\frac{\partial }{\partial r}m\left(v,r\right)$ and 
 $\dot{m}=\frac{\partial }{\partial v} m\left(v,r\right)$.

For our solution, we propose a two-fluid energy-momentum tensor comprising a timelike static fluid with a timelike four-velocity given by $u^{a}=\left(\frac{1}{\sqrt{F}},\,0,\,0,\,0\right)$ and a two-component null fluid with the four-velocities $v^{a}=\left(0,\,1,\,0,\,0\right)$ and $w^{a}=\left(1,\,-\frac{F}{2},\,0,\,0\right)$. The massive or timelike matter part of the energy-momentum tensor can be expressed as:
\begin{equation}
T_{\:b}^{a}=(\rho_{m}+p_{m})u^{a}u_{b}-p_{m}\delta_{\:b}^{a},
\label{eq:radtab}
\end{equation}
 With  eq.~(\ref{eq:eint:radiation}), we can solve for $\rho_m$ and $p_m$ can be written as,
\begin{eqnarray}
\rho_{m} & = & \frac{3}{R^{2}}F\left(v,r\right)\, , \nonumber \\
p_{m} & = & -\frac{1}{R^{2}}F\left(v,r\right)\,.
\end{eqnarray}
 We observe that the timelike matter field still satisfies the equation of state $\rho_{m}+3p_{m}=0,$ indicating special cases of Whittaker solutions. We use the  form of the energy-momentum tensor for the null fluid given  by  Husain~\cite{VH1996}.
\begin{equation}
\bar{T}_{\:b}^{a}=-\frac{2}{R^{2}}\dot{m}\left(v,r\right)\,\cot^{2}\left(\frac{r}{R}\right)v^{a}v_{b}+\rho_{r}\left(v^{a}w_{b}+w^{a}v_{b}\right)+p_{r}\left(-\delta_{\:b}^{a}+v^{a}w_{b}+w^{a}v_{b}\right)
\end{equation}
 where,
\begin{eqnarray*}
\rho_{r} & = & \frac{2}{R^{2}}m'\left(v,r\right)\cot^{2}\left(\frac{r}{R}\right)\,,\\
p_{r} & = & \frac{2}{R^{2}}m'\left(v,r\right)-\frac{1}{R}\cot\left(\frac{r}{R}\right)m''\left(v,r\right)
\end{eqnarray*}
 By selecting a suitable equation of state for $\left(\rho_{r},\,p_{r}\right)$, one can derive an explicit solution for $m\left(v,r\right)$\cite{VH1996, Maharaj2021}. We draw attention to a straightforward special case of a Vaidya-like radiating solution by setting $m'\left(v,r\right)=m''\left(v,r\right)=0$, to obtain the energy-momentum tensor with
\begin{equation}
\bar{T}_{\:b}^{a}=-\frac{2}{R^{2}}\dot{m}\left(v\right)\,\cot^{2}\left(\frac{r}{R}\right)v^{a}v_{b}\,,
\end{equation}
For the metric given in eq.~(\ref{eq:radds}), as $m$ is an arbitrary function of $t-r$, the geometry lacks time symmetry, i.e., there is no timelike Killing-vector. The blackhole can shrink or grow depending on the flow of radiation inward or outward. This demonstrates the possibility of the existence of several interesting asymptotically non-flat blackhole solutions. These solutions can play an important role in investigating the properties of 
blackholes in non-flat backgrounds.
\begin{figure}[t]
\begin{center}
\subfloat[]{
\includegraphics[scale=0.29]{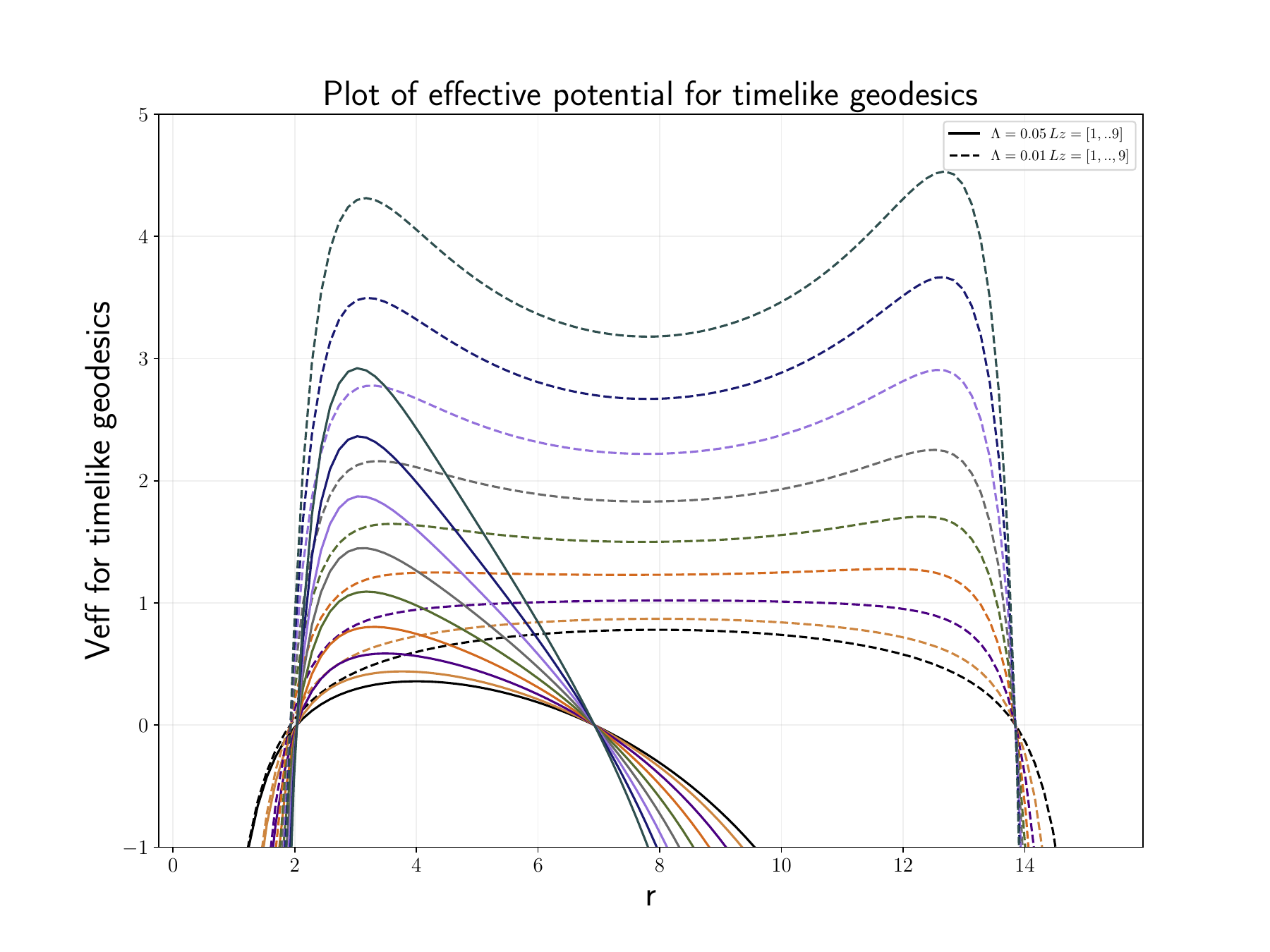} 
\label{eq:timelikepot} 
}
\subfloat[]{
\includegraphics[scale=0.29]{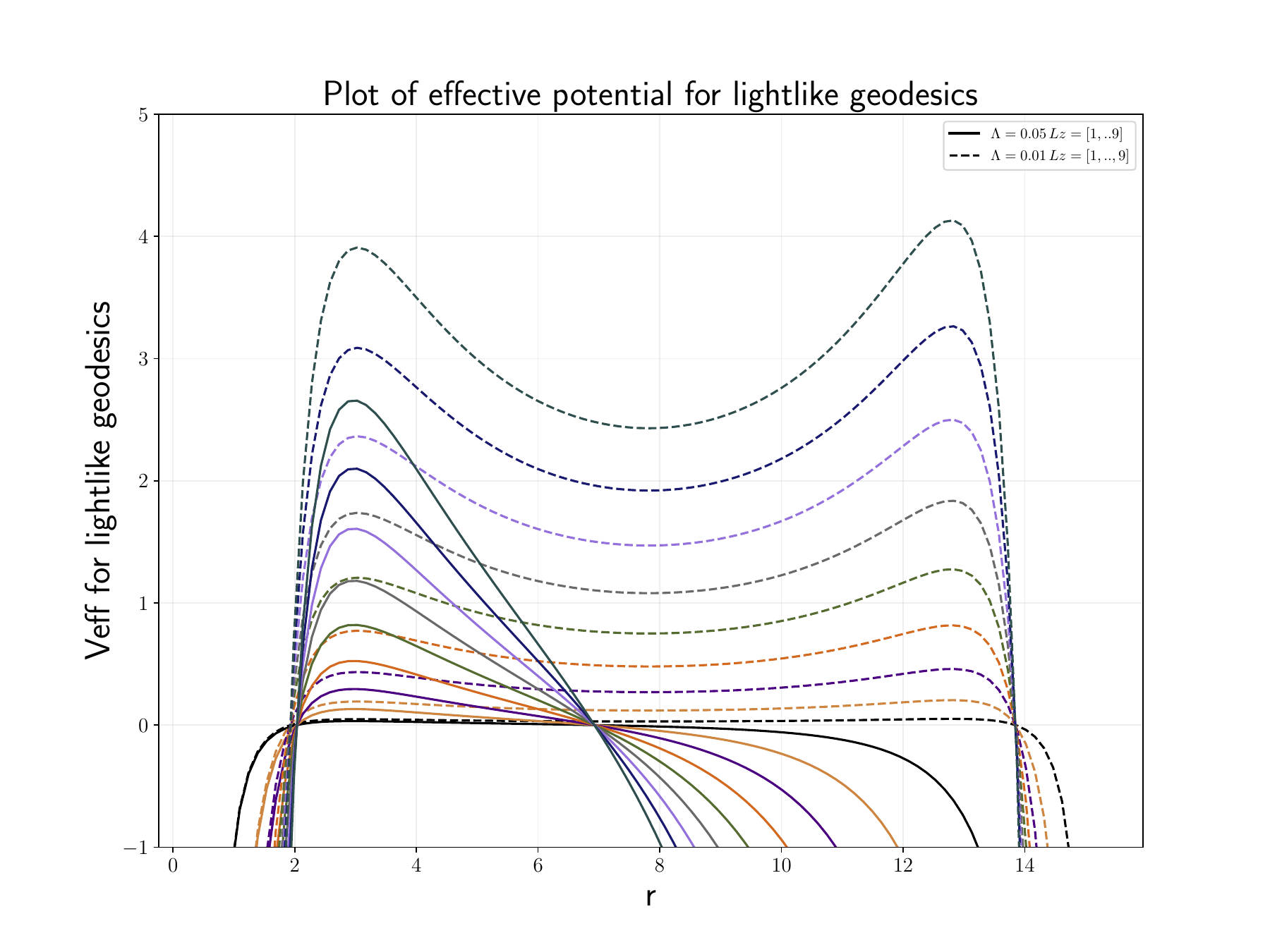}
\label{eq:lightlikepot} 
}
\caption{\ \\(a): Plot of the effective potential, $V_{\mathrm{eff}}$, for timelike geodesics as a function of $r$. We fix the value of $R=5$, while using two representative values of $\Lambda$, i.e., $\Lambda=0.05$ with dashed curves and $\Lambda=0.01$ with solid lines.
\\
(b): Plot of the effective potential, $V_{\mathrm{eff}}$, for lightlike geodesics as a function of $r$. We fix the value of $R=5$, while using two representative values of $\Lambda$, i.e., $\Lambda=0.05$ with dashed curves and $\Lambda=0.01$ with solid lines. }
\end{center}
\end{figure}
\section{Event Horizon and Static Observer\label{sec:EHand_SO}}
The Killing-horizon in the static spherically symmetric spacetime is determined by the condition $g_{00}=0$. For the SW spacetime, this condition yields the explicit expression
\begin{equation}
R \tan\left(\frac rR\right)\left(1-\Lambda R^{2}\right)+\left(\Lambda R^2 r-2m\right)=0\,.
\label{eq:condition_for_horizon}
\end{equation}
In general, there are two horizons: one blackhole horizon closer to the center and the other cosmological horizon closer to the outer boundary. Since it is difficult to provide an explicit analytical expression for the position of these horizons,
 we numerically determine the horizon's location using the condition given by eq.~(\ref{eq:condition_for_horizon}). First, we find the horizon's location by fixing a range of values of $\Lambda$ from $0.001$ to $0.25$ at equal intervals and plot them as a function of $R$. In Figure-\ref{fig:Horizon_funR}, the solid lines represent the cosmological horizon, while the dashed lines represent the blackhole event horizon. For smaller values of $\Lambda$,
the solution approaches Vaidya's static solution, and for larger values of $R$, it matches with the SdS solution. Finally, for some combinations of values of $R$ and $\Lambda$, both horizons meet, and no part of the Universe has a static timelike observer. This makes the fluid-based source model with the Whittaker equation of state fail. We refer to this as an extremal condition, as shown in Figure-\ref{fig:Horizon_funR}. Such possibilities are not new; similar situations are observed in SdS and also blackhole in an expanding Universe~\cite{PhysRevD.78.024008}.

We also plot the location of horizons by reversing the roles of $\Lambda$ and $R$. In this case, we fix a set of values of $R$ from $1.58$ to $12.0$ at equal intervals and plot the location of the horizon as a function of $\Lambda$,
as shown in Figure-\ref{fig:Horizon_funLam}. From these studies, we conclude that the larger the values of $R$, the larger both horizons become, and as $R \rightarrow \infty$, we have the SdS case. However, the blackhole horizon seems to increase in size with increasing $R$, while the cosmological horizon appears to shrink with an increase in $\Lambda$. 

Because SW spacetime is a static solution, the event horizon and the static limit match. In usual cases, such as the Schwarzschild solution, a static observer becomes geodesic only at spatial infinity. However, in SW blackholes, due to contributions from $M$, $R$, and $\Lambda$, a static observer experiences both inward and outward pulls, or attractions in the direction of both horizons. At certain locations, these forces can balance, allowing a static observer to follow geodesic motion at a finite distance. The acceleration of a static observer is given by,
\begin{equation}
a_{q}= \phi_{,q} \, \, \,  \mathrm{where} \, \,  e^{-2\phi}=g_{00}\,.
\end{equation}
The direction of the acceleration is toward the horizons. In between the two horizons, the acceleration becomes zero, and a static observer follows a geodesic motion. The spatial location at which the acceleration becomes zero, allowing a static observer to follow a geodesic motion, is shown in 
Figure-\ref{fig:static_geo}.
Here also, we first fix values of $\Lambda$ and solve for the location of the geodesic as a function of $R$,
as shown in Figure-\ref{fig:static_geo}-(a), we observe that an increase in $\Lambda$ brings the location of the geodesic closer to the blackhole horizon. An increase in $R$ pushes the location further from the blackhole horizon. In Figure-\ref{fig:static_geo}-(b), we reverse the roles of $\Lambda$ and $R$ and plot the location of a static observer as a function of $\Lambda$ for a fixed value of $R$.
Increasing $\Lambda$ pushes the geodesic position of a static particle toward the cosmological horizons. 
\begin{figure}[t]
\begin{center}
\begin{minipage}{0.35\textwidth}
\subfloat[Orit Intergration]{%
	\includegraphics[scale=0.11]{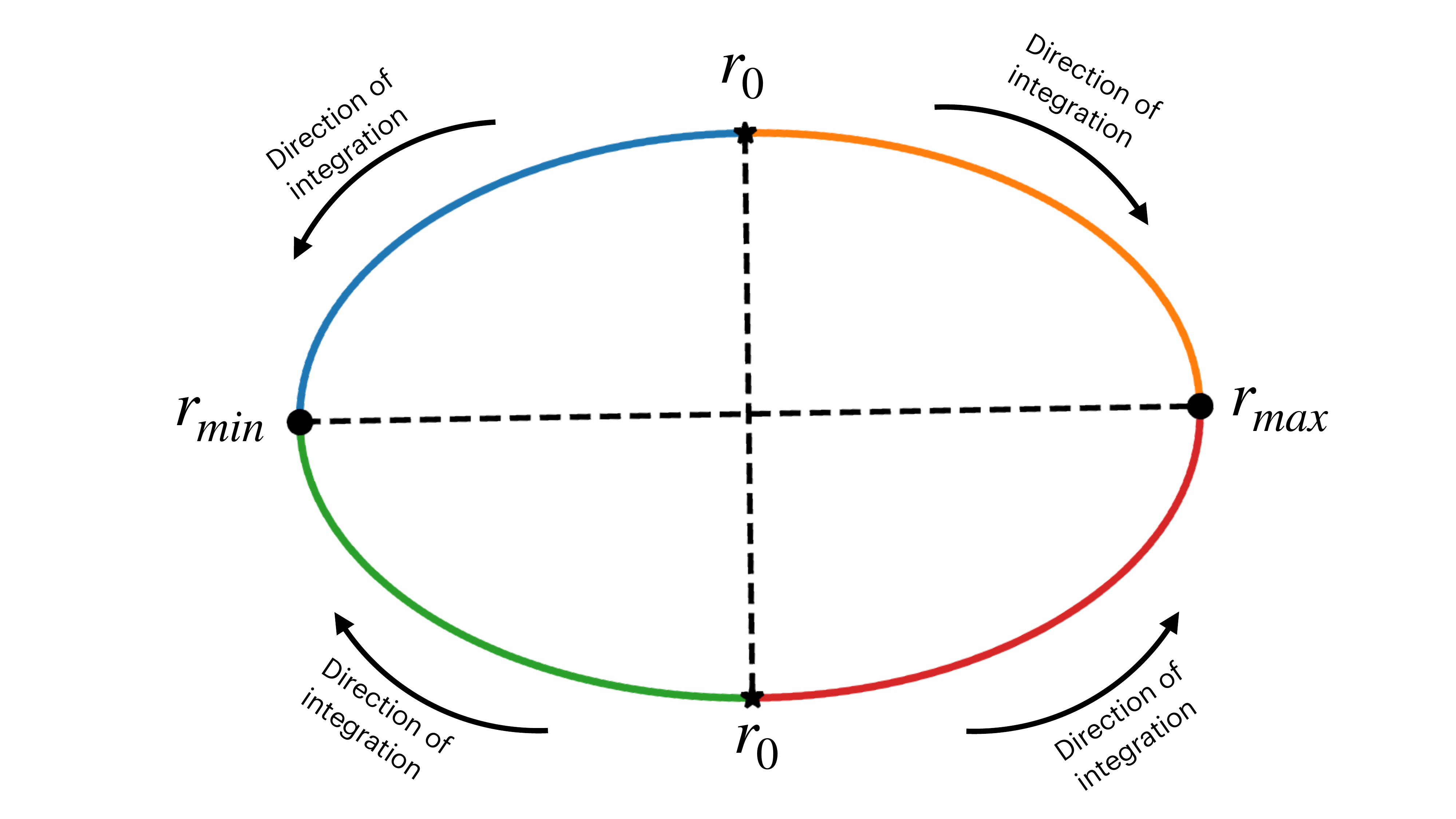}%
        \label{fig:orbit_int}%
        }
 \end{minipage}
 \begin{minipage}{0.5\textwidth}
    \subfloat[Newtonian Orbit]{%
        \includegraphics[scale=0.3]{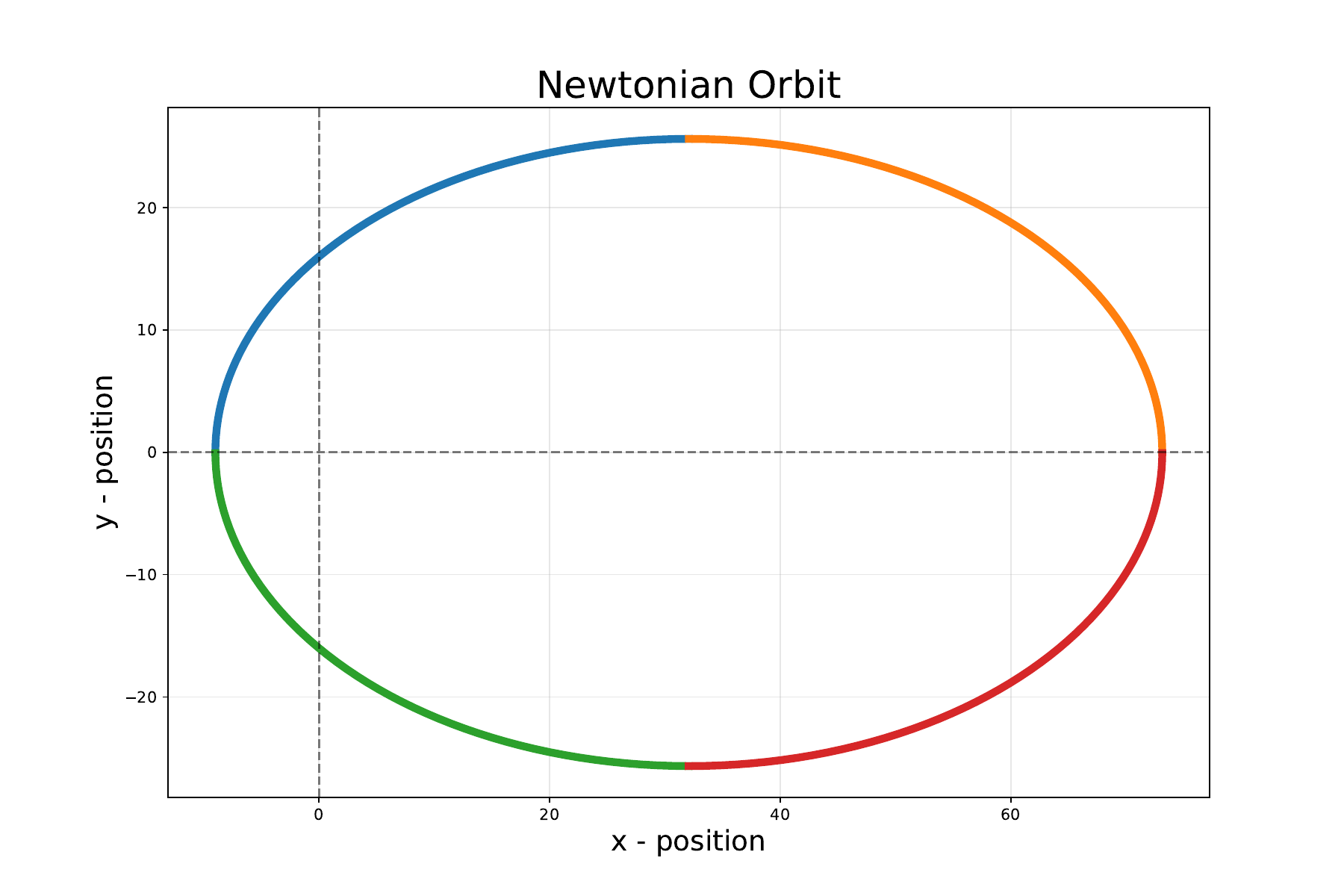}%
        \label{fig:Newtonian}%
        }%
\end{minipage}
\caption{The orbital equation is integrated by combining four sections. We use $r_0 = (r_{min}+r_{max})/2$ as the radial starting point, where $r_{min}$ and $r_{max}$ are the radial turning points for a given $Lz$ and $E_t$. The first two sections of orbits are obtained by integrating the orbital equation, starting from $\{r_0, \pi/2\}$ and proceeding inward ($\epsilon=-1$) and outward ($\epsilon=1$). Similarly, the other two sections are obtained by starting from $\{r_0, 3\pi/2\}$ and integrating in the inward and outward directions. We combine all four sections smoothly to obtain the complete orbit. 
\label{fig:orbit_direction}}
\end{center}
\end{figure}
\section{Geodesic motion and Periastron Precession \label{sec:geo_prec}}
In this section, we investigate the geodesic motion in SW spacetime. Since the metric given in eq.(\ref{eq:ds2}) is static and spherically symmetric, it is straightforward to obtain the geodesic equation of motion. Additionally, we restrict ourselves to the equatorial plane, i.e., $\theta=\frac{\pi}{2}$, $\dot{\theta}=0$, and $\ddot{\theta}=0$. In this case, the timelike geodesic equation can be easily separated using constants of motion. With these, we obtain:
\begin{eqnarray}
\dot{t} & = & \frac{E_{t}}{F(r)}\,,\nonumber\nonumber \\
\dot{\phi} & = & \frac{L_{Z}}{R^2 \sin^2\left(\frac{r}{R}\right)}\,,\nonumber\nonumber \\
\dot{r}^{2} & = & E_{t}^{2}-F(r)\left[1+\frac{L_{z}^{2}}{R^2 \sin^2\left(\frac{r}{R}\right)}\right]\,.\label{eq:geod_timelike}
\end{eqnarray}
Here, $E_{t}$ and $L_{z}$ are constants of motion, namely, the energy and the z-component of angular momentum for the timelike particle. For the light or null geodesic, the corresponding equations are given by:
\begin{eqnarray}
\dot{t} & = & \frac{E_{p}}{F(r)}\,,\nonumber\nonumber \\
\dot{\phi} & = & \frac{L_{p}}{R^2 \sin^2\left(\frac{r}{R}\right)}\,,\nonumber\nonumber \\
\dot{r}^{2} & = & E_{p}^{2}-F(r)\frac{L_{p}^{2}}{R^2 \sin^2\left(\frac{r}{R}\right)}\,.\label{eq:geod_null}
\end{eqnarray}
In this context, we also have $E_p$ and $L_p$ as conserved quantities. From the radial equation, we can derive the effective potential for timelike and null geodesics. The potentials are given by:
\begin{eqnarray}
V_{\mathrm{eff}}&=&F(r) \left[1\,+\,\frac{L_{z}^{2}}{R^2 \sin^2\left(\frac{r}{R}\right)}\right]\,, \nonumber \\
V_{\mathrm{null}}&=& F(r)\frac{L_{p}^{2}}{R^2 \sin^2\left(\frac{r}{R}\right)}\,.
\end{eqnarray}
With the function $F(r)$ given by equation (\ref{eq:g00}), the maxima and minima of the effective potential correspond to unstable and stable circular geodesics. An important difference in the SdS solution, the position of circular null geodesics does not depend on $\Lambda$. However, in the case of SW spacetime, the potential is not just an additive constant, and the position of null geodesics explicitly depends on $\Lambda$. The effective potentials for both timelike and null geodesics exhibit similar behaviour, as shown in Figures \ref{eq:timelikepot} and \ref{eq:lightlikepot}. 

For smaller values of $\Lambda$, there exists a single unstable circular geodesic. In contrast, for larger values of $\Lambda$, the number and nature of circular geodesics depend significantly on the value of $L_z$. As illustrated in the figures, a low value of $L_z$ results in a single unstable circular geodesic. When $L_z=0$, it corresponds to a static observer, as previously discussed. Increasing the value of $L_z$ introduces a stable circular geodesic in the middle, with two unstable circular geodesics appearing closer to each horizon. This variation is indeed an intriguing difference.

%
%
\begin{figure}[t]
\begin{center}
\includegraphics[scale=0.5]{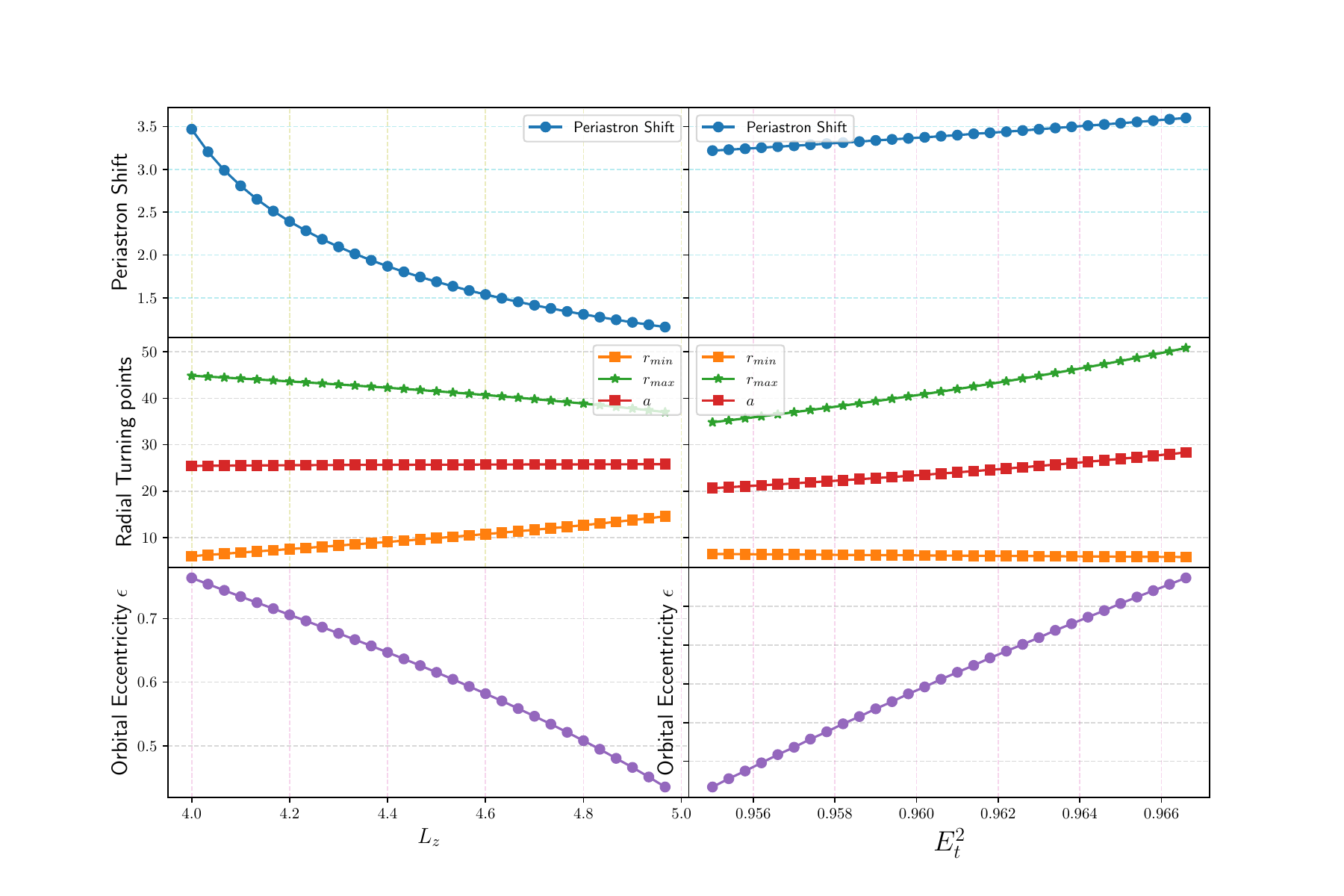}
\caption{ Plots of periastron shift and radial turning points for a massive particle in Schwarzschild spacetime as functions of $L_z$ and ${E_t}^2$ for $M=1$. The plots in the top row show the behavior of the periastron, while those in the bottom row display the behavior of the radial turning points $r_{min}$ and $r_{max}$. 
 \label{fig:sch_periastron}}
\end{center}
\end{figure}
\begin{figure}[t]
\begin{center}
\includegraphics[scale=0.5]{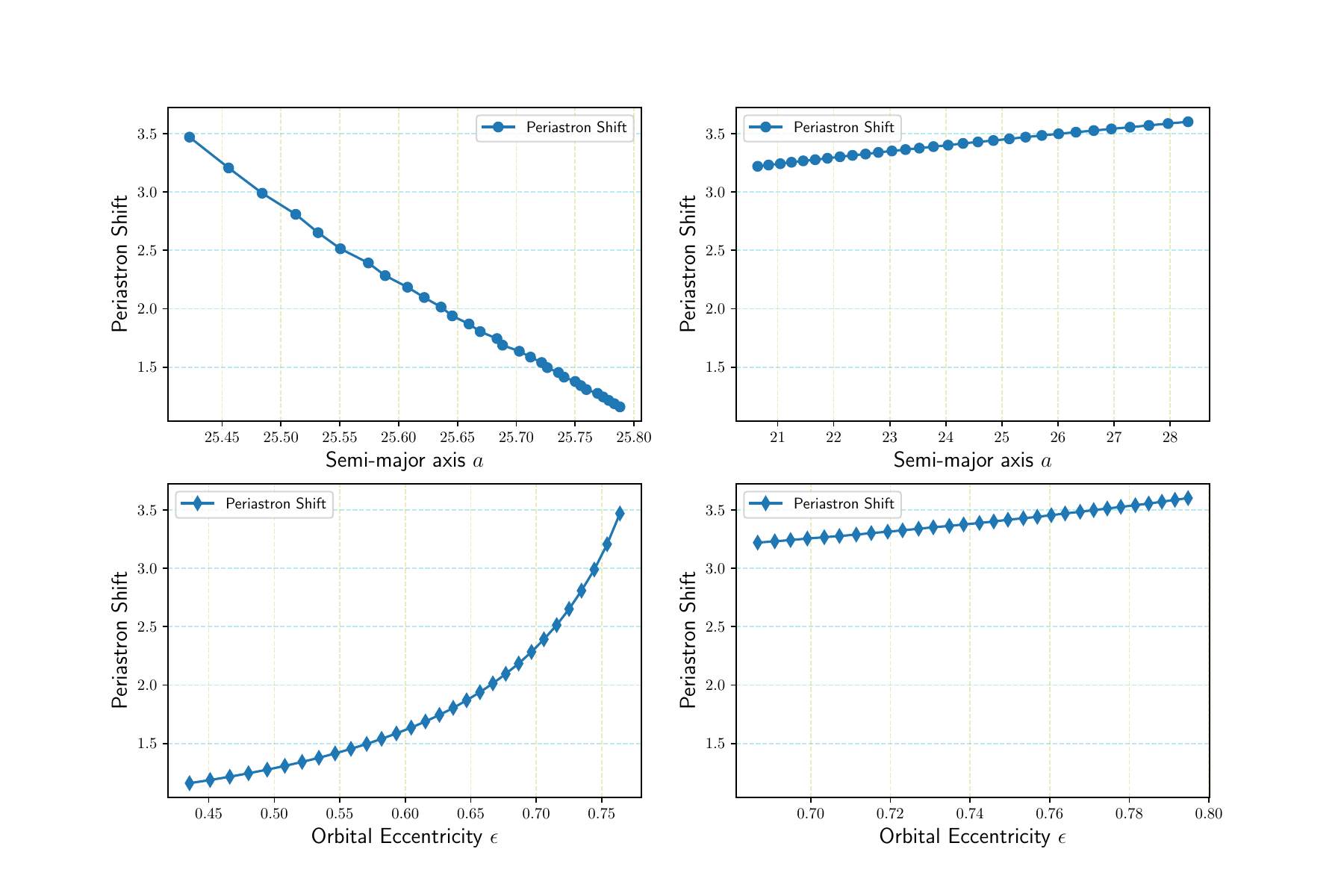}
\caption{Plots of periastron shift as functions of the semi-major axis, $a$, and orbital eccentricity, $\epsilon$, in the Schwarzschild spacetime are presented. The plots in the left column correspond to variations in $L_z$ and depict the precession behavior as a function of $a$ and $\epsilon$. Similarly, the plots in the right column are obtained by varying the energy parameter, $E_t$.
 \label{fig:sch_periastronb}}
\end{center}
\end{figure} 
 \subsection{Precession of Periastron} 
In this section, we examine the precession of the periastron for timelike elliptical orbits. We begin by utilizing the timelike geodesic equations provided in eq.~(\ref{eq:geod_timelike}), and we numerically integrate the $r,\,\phi$ equations to obtain the complete orbit. In this case, we integrate two first-order equations, as follows:
\begin{eqnarray}
\frac{d\phi}{d\tau} & = & \frac{L_{z}}{R^2 \sin^2\left(\frac{r}{R}\right)}\,,\nonumber \\
\frac{dr}{d\tau} & = &  \epsilon  \left[ E_{t}^{2}-F(r)-F(r)\frac{L_{z}^{2}}{R^2 \sin^2\left(\frac{r}{R}\right)}\right]^{1/2}\,.
\label{eq:orbit_eq_wbh}
\end{eqnarray}
In the case of the radial equation, the parameter $\epsilon$ represents the direction of propagation, i.e., when $\epsilon = +1$, the particle moves along the outwards or increasing direction of $r$. On the other hand, when $\epsilon = -1$, the radial coordinate decreases, causing the particle to move inwards. We integrate the $r$ and $\phi$ equations as functions of proper time $\tau$, and this allows us to easily construct the spatial orbital equation $r(\phi)$. For completeness and comparison, we also explicitly provide the time-like geodesic equation for the Schwarzschild case.
\begin{eqnarray}
\frac{d\phi}{d\tau} & = & \frac{L_{z}}{r^2}\,,\nonumber \\
\frac{dr}{d\tau} & = &  \epsilon  \left[ E_{t}^{2}-1+\frac{2M}{r}-\frac{{L_z}^2}{r^2}+ \frac{2 M \, {L_z}^2}{r^3 }\right]^{1/2}\,,
\label{eq:orbit_eq_sch}
\end{eqnarray}
and in  Schwarzschild-de Sitter case,
\begin{eqnarray}
\frac{d\phi}{d\tau} & = & \frac{L_{z}}{r^2}\,,\nonumber \\
\frac{dr}{d\tau} & = &  \epsilon  \left[ E_{t}^{2}-\left(1-\frac{2M}{r}-\frac{\Lambda}{3} r^3\right)\left(1+\frac{{L_z}^2}{r^2}\right)\right]^{1/2}\,.
\label{eq:orbit_eq_schde}
\end{eqnarray}
However, the radial turning points correspond to stationary points in the equation of motion at which the differential equation breaks down and poses difficulties in the numerical integration, leading to large numerical errors. 
To avoid these difficulties, we break the integration of orbital differential equations into four sections, with each section ending at a stationary point. We smoothly combine the solutions to obtain the full orbit equation, as shown in Figure-\ref{fig:orbit_int}. We use the following simple steps to achieve accurate results using the Euler method to integrate the $\phi$ and $r$ equations. As a test case for the procedure, in the case of a Newtonian potential or Kepler problem, we obtain an exact elliptical orbit without any precession. However, a deviation from the Newtonian potential leads to a more complicated orbit, which can be assumed to be instantaneously elliptical, with its semi-major axis precessing over time. Due to the time-independent or static nature of the spacetime, $r_{min}$ and $r_{max}$ remain the same as functions of time. However, they slowly drift as functions of $\phi$, resulting in a slow drift in the orientation of the semi-major axis. It is this shift or change in the semi-major axis that we measure and serves as one of the first tests for general relativity. Our interest is not in testing the theory of gravity; rather, we aim to evaluate the influence of cosmological parameters, such as $R$ and $\Lambda$, on the phenomenon of periastron precession. We measure the periastron precession closer to the event horizon, or in the strong field limits. For this reason, we do not use the perturbative methods predominantly  used in the similar studies,  
instead we numerically integrate the orbit equation. Once again, our interest is not to compare deviations from Newtonian theory or measure small changes in astrophysical scenarios. We compare deviations with respect to a Schwarzschild solutions and blackhole  influenced by cosmological parameters.

\begin{figure}[t]
\begin{center}
\includegraphics[scale=0.5]{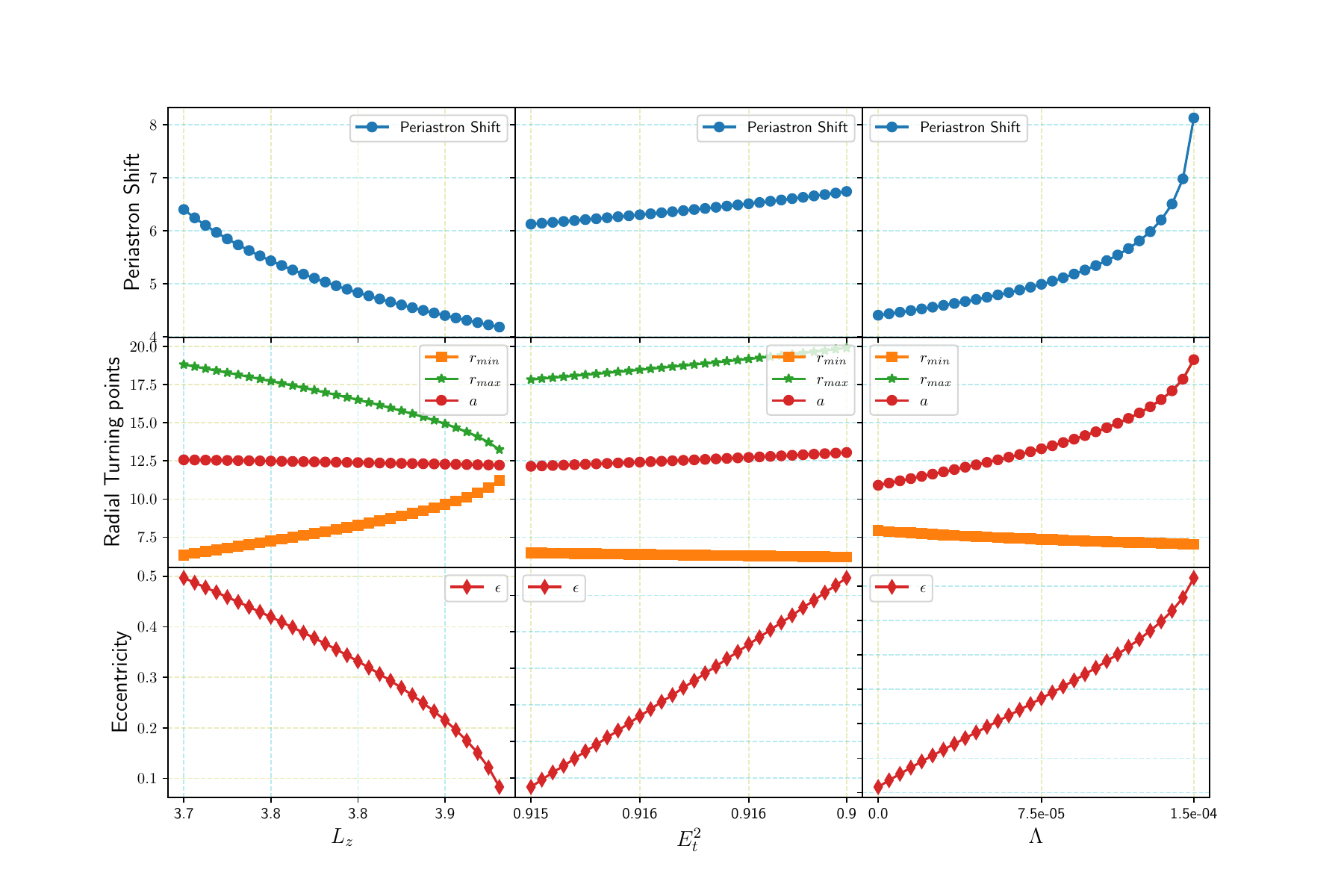}
\caption{Plots of periastron shift and radial turning points for a massive particle in the SdS spacetime as functions of various parameters. The plots in the top row show the dependence of periastron on the parameters $Lz$, $E_t^2$, and $\Lambda$, respectively. Similarly, the bottom row shows the dependence of radial turning points ($r_min$ and $r_max$) as functions of $Lz$, $E_t^2$, and $Lambda$. The value of mass is fixed at $M=1$.  \label{fig:sde_periastron}}
\end{center}
\end{figure}
\begin{figure}[t]
\begin{center}
\includegraphics[scale=0.5]{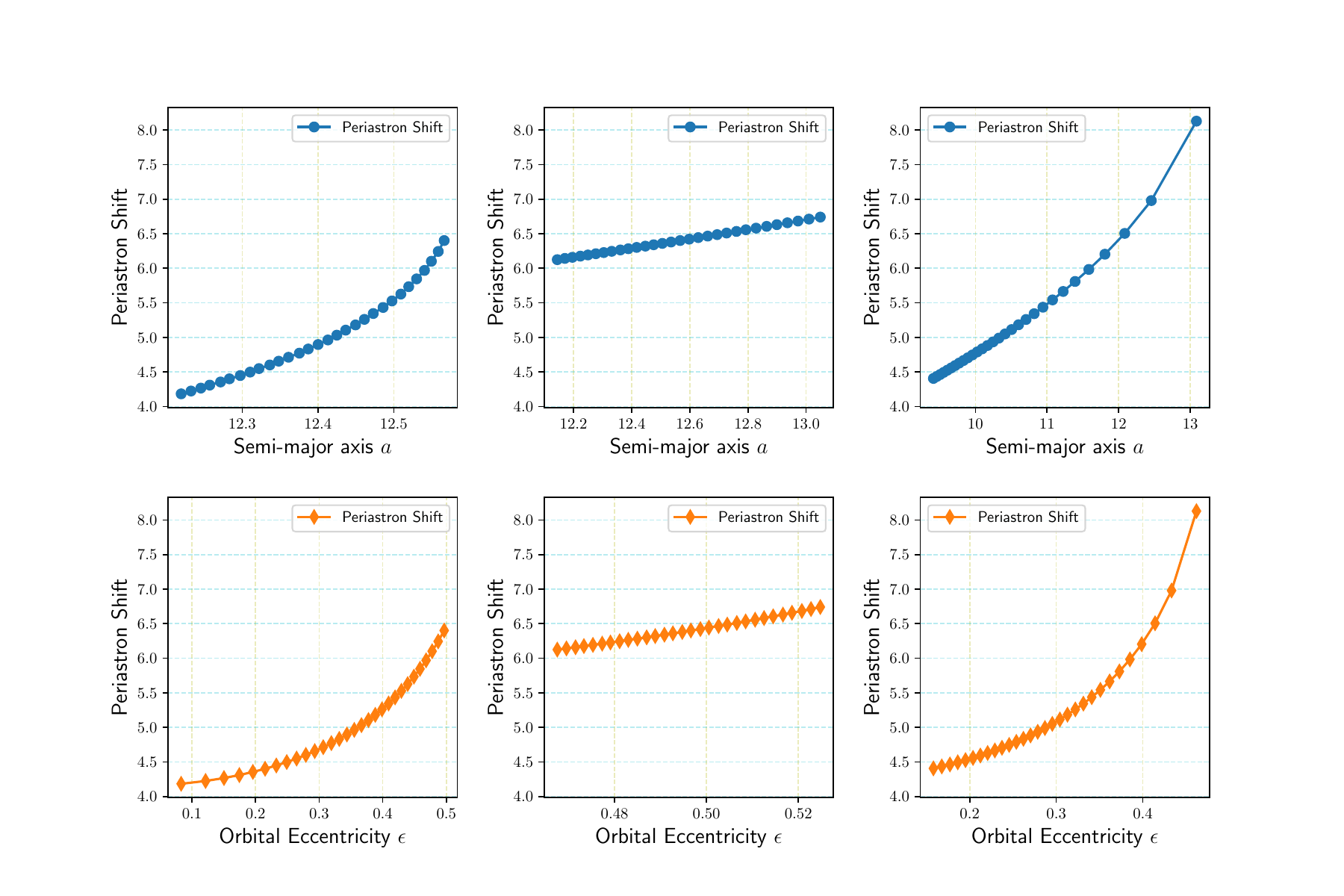}
\caption{ Plots of precession of periastron as a function of semi-major axis ($a$) and eccentricity ($\epsilon$) in the SdS spacetime. In the rightmost column, variations in ($a$) and ($\epsilon$) are induced by varying $L_z$. Similarly, in the second and third columns, changes in ($a$) and ($\epsilon$) are induced by varying $E_t$ and $\Lambda$, respectively.  
  \label{fig:sde_periastronb}}
\end{center}
\end{figure}
Numerical  estimating precession involves the following steps:
\begin{enumerate}
\item For a given $E_t$ and $L_z$, we determine the radial points $r_{min}$ and $r_{max}$. In the case of an elliptical orbit, the length of the semi-major axis $a$ is given by,
$a = \frac{1}{2}(r_{min} + r_{max})$.
The length of the semi-minor axis $b$ is calculated as,
$b = \sqrt{r_{min} \cdot r_{max}}$.
The eccentricity, denoted as $\epsilon$, can be computed as:
$\epsilon = \sqrt{1 - \frac{b^2}{a^2}}$.
Due to stationary points at $r_{min}$ and $r_{max}$, integrating the orbital ordinary differential equations in one go can result in large errors. To minimize numerical errors, we divide the orbit into four sections. Starting from the initial coordinates $\{ r = r_0, \, \phi = \frac{\pi}{2}\}$, we first integrate along the increasing $r$ direction ($\epsilon = +1$) until we reach $r_{max}$. Then, to obtain the orbit from $\{ r_0, \, \frac{\pi}{2}\}$ to $r_{min}$, we integrate along the decreasing $r$ direction ($\epsilon = -1$) until we reach $r_{min}$. We use a similar approach to integrate the orbital equations starting from $\{ r = r_0, \, \phi = \frac{3\pi}{2}\}$ and arriving at $r_{max}$ and $r_{min}$. The details are illustrated in Figure~\ref{fig:orbit_int}.

\item We smoothly combine all four patches to construct the complete orbit. For testing, we use the Newtonian potential for which the orbit is elliptical with no periastron precession, as shown in figure \ref{fig:Newtonian}. The value of the periastron precession, denoted as $\phi(\tau_0)-2\pi$, is calculated, where $\tau_0$ is the proper time taken to complete the orbit from $r_{max} \rightarrow r_{min} \rightarrow r_{max}$.
\end{enumerate}
We study the SW, SdS, and Schwarzschild solutions separately, with the Schwarzschild case serving as a reference. We investigate the influence of $\Lambda$ on the de-Sitter case and the influence of $\Lambda$ and $R$ on SW, considering the variations in orbital parameters such as $E_t$ and $L_z$, which in turn affect $r_{min}$ and $r_{max}$.
\begin{figure}[t]
\begin{center}
\includegraphics[scale=0.5]{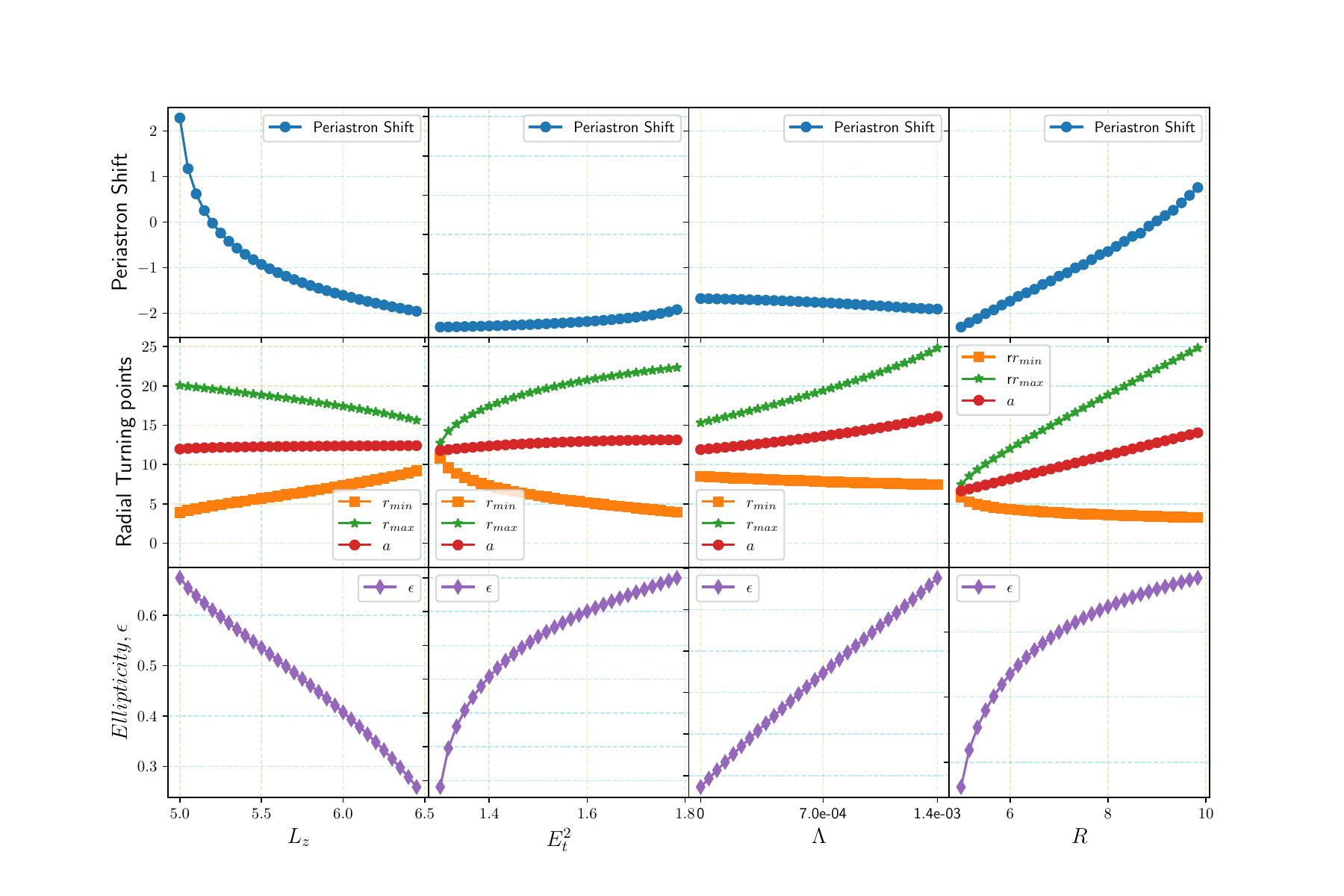}
\caption{ Plots of periastron shift and radial turning points for a massive particle in SW spacetime are presented as functions of various parameters. In this case, we fix the value of the mass, $M=1$. The top row of plots depicts the dependence of periastron on various parameters, including $L_z, \,{E_t}^2, \, \Lambda, \, \rm{and} , \, R$. Similarly, the bottom row shows the dependence of radial turning points, $r_{min}$ and $r_{max}$, as functions of $L_z,\, {E_t}^2, \,\Lambda, \,\rm{and} , \, R,$  respectively.
 \label{fig:wbh_periastron} } 
\end{center}
\end{figure} 
\begin{figure}[t]
\begin{center}
\includegraphics[scale=0.5]{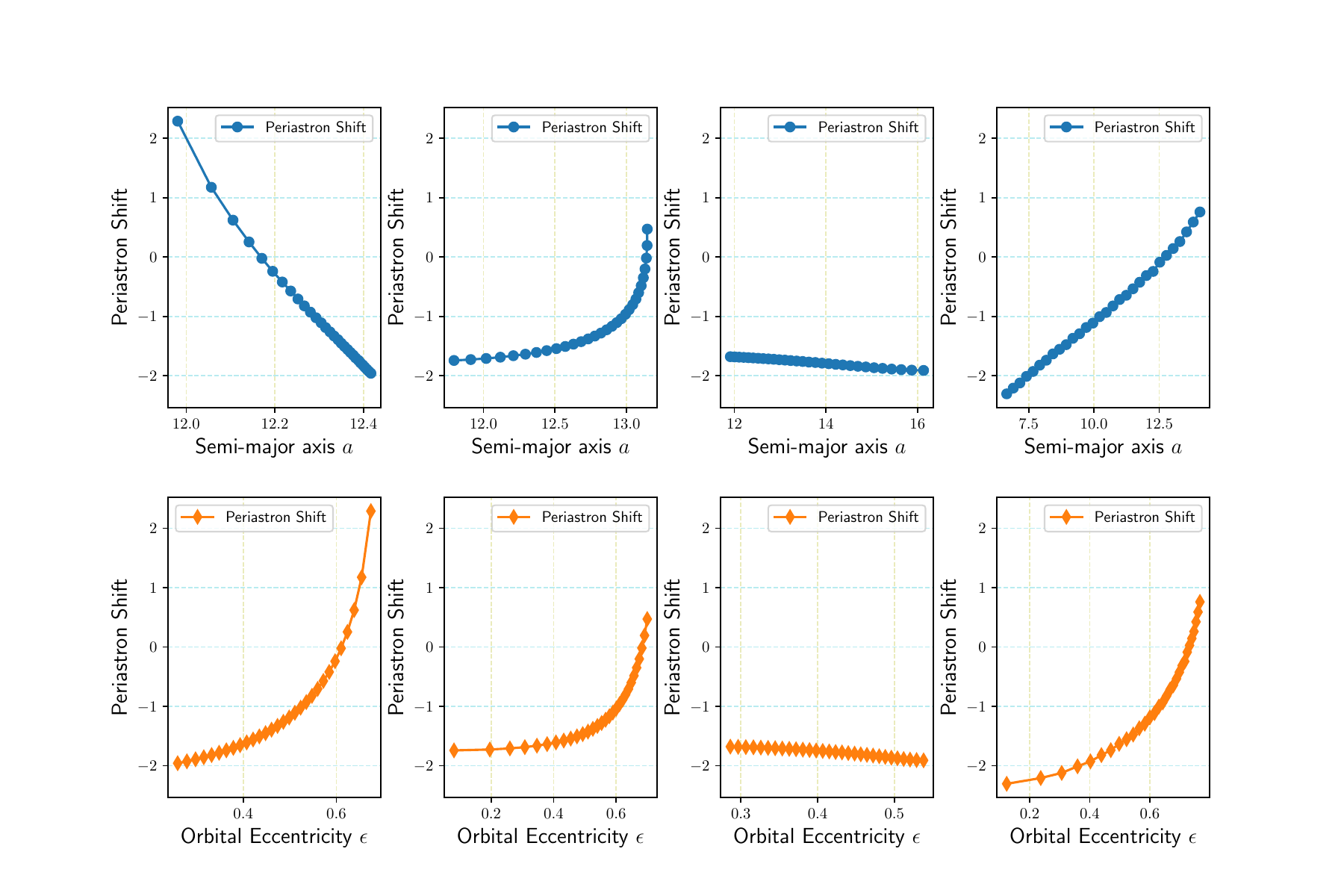}
\caption{Plots of precession of periastron as a function of semi-major axis, $a$, and eccentricity, $\epsilon$, in SW spacetime are presented. In the rightmost column, the changes in $a$ and $\epsilon$ are induced by varying $L_z$. Similarly, in the next three columns, the changes in $a$ and $\epsilon$ are induced by varying $E_t$, $\Lambda$, and $R$, respectively.
 \label{fig:wbh_periastronb} } 
\end{center}
\end{figure} 
\subsubsection{Schwarzschild Spacetime}
First, we look at the precession of periastron in the case of the Schwarzschild blackhole as our reference. We arbitrarily choose a range of $E_t$ and $L_z$ such that $r_{min}$ is sufficiently close to the event horizon but not too close. We start by fixing the value of $E_t^2$ and find ${r_{min},\, r_{max}}$ by varying $L_z$. Once we have obtained the turning points, we integrate the orbital equation to calculate the periastron precession. The results are summarized in figure-\ref{fig:sch_periastron}. It is easy to see that the precession is sensitive to $L_z$, as it determines the shape of the effective potential. However, the precession weakly depends on ${E_t}^2$.
The orbital eccentricity changes considerably within the range of $L_z$ values considered. In Figure \ref{fig:sch_periastron}, we observe a linear dependence of periastron precession on the semi-major axis, while its relationship with orbital eccentricity is more complex. Figures \ref{fig:sch_periastron} and \ref{fig:sch_periastronb} reveal a weak dependence on $E_t$. Although eccentricity increases with higher $E_t$ values, leading to an increase in the semi-major axis, the combined effect results in only a marginal change in periastron precession.
\subsubsection{Schwarzschild de-Sitter Spacetime}
In this section, we explore the influence of the cosmological constant on the precession of the periastron using the SdS solution. Our aim is not to constrain the cosmological constant in weaker gravitational fields such as the Sun but to examine the qualitative differences in the strong field due to the presence of the cosmological constant and understand the dependence of precession on the value of $\Lambda$. In this case, we have three parameters: $L_z^2$, $E_t$, and $\Lambda$. As before, we fix any two parameters within the given range and vary the third parameter to determine the turning points. With the turning points, we can integrate the orbital equation to determine the periastron shift. We repeat this procedure to obtain the periastron as a function of all parameters. The results are shown in Figure \ref{fig:sde_periastron}. Although $\Lambda$ has to be reasonably small, it has a considerable influence on the periastron precession. Similar to the case of the Schwarzschild blackhole, in this case, the precession weakly depends on the energy of the particle. In Figure \ref{fig:sde_periastronb}, we observe that in the case of SdS spacetime, the dependence on the semi-major axis obtained by changing $L_z$ is stronger than in the Schwarzschild case. Meanwhile, the dependency with eccentricity is very similar. In the case of SdS spacetime, the dependency of periastron precession on orbital energy is weak.
\begin{figure}[t]
\begin{center}
\includegraphics[scale=0.6]{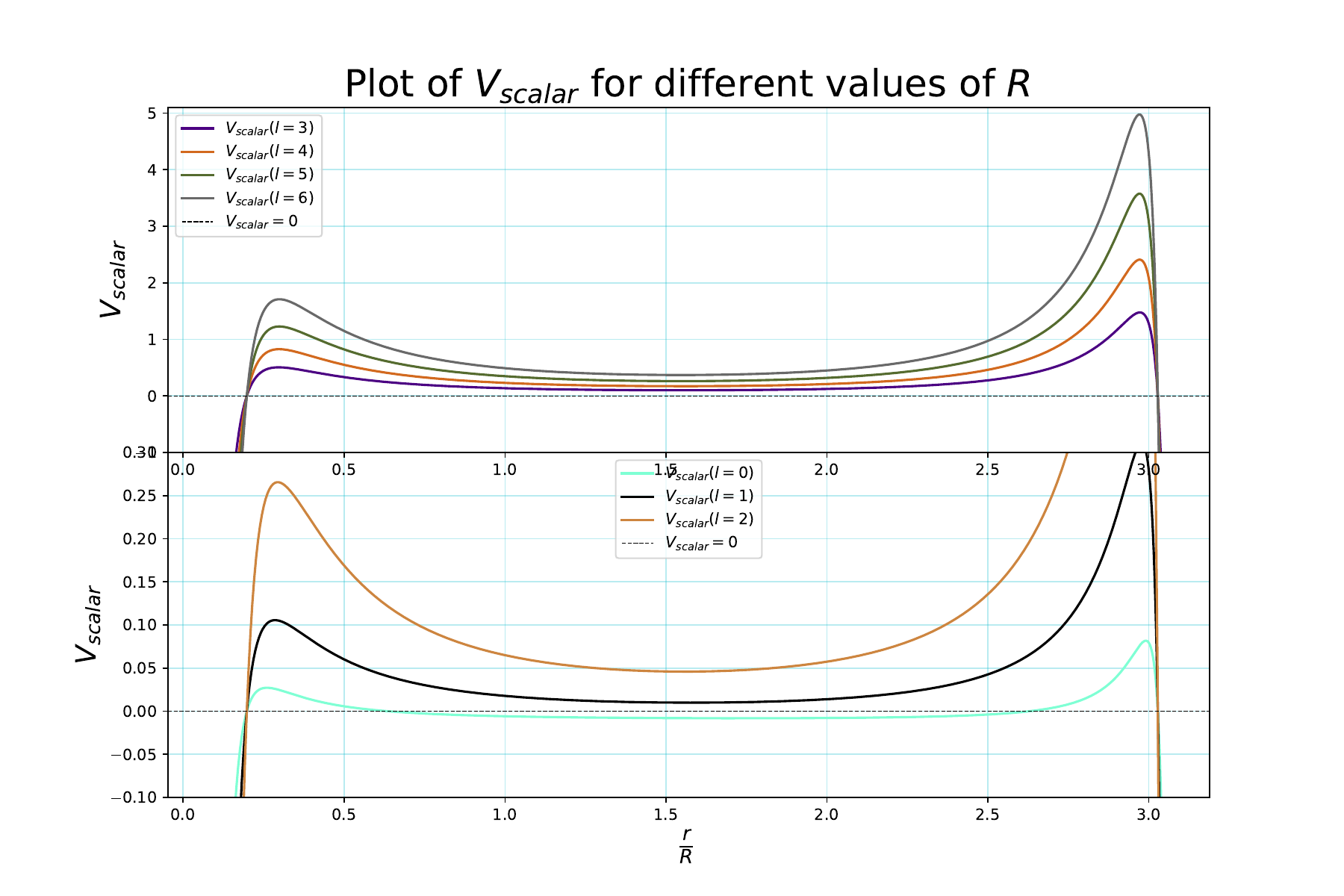}
\caption{Plots of the effective potential for a scalar field as a function of $\frac{r}{R}$ for different values of $l$ ranging from 0 to 6. We have fixed the values of $R=10$, $M=1$, and $\Lambda=0.001$.
\label{fig:VR1}}
\end{center}
\end{figure}

\subsubsection{Schwarzschild-Whittaker Spacetime}
In this section, we investigate the periastron precession of elliptical orbits in the SW blackhole, which is influenced by two parameters, $R$ and $\Lambda$, introducing cosmological effects. This results in a total of four parameters: $L_z$, $E_t$, $R$, and $\Lambda$. We systematically study the precession of the periastron, similar to our approach in the previous case. The results are summarized in Figure-\ref{fig:wbh_periastron}. Interestingly, it appears that the contribution to precession from $R$ tends to suppress the effects of $\Lambda$. However, this effect may be attributed to the fact that $R$ has a larger impact on the orbital parameters $r_{min}$ and $r_{max}$. While precession is directly dependent on $r_{min}$ and $r_{max}$, we have yet to determine the precise independent relationship between the two.
\begin{figure}[t]
\begin{center}
\subfloat[$V_{scalar}(r)$ for $l=0$ ]{%
	\includegraphics[scale=0.3]{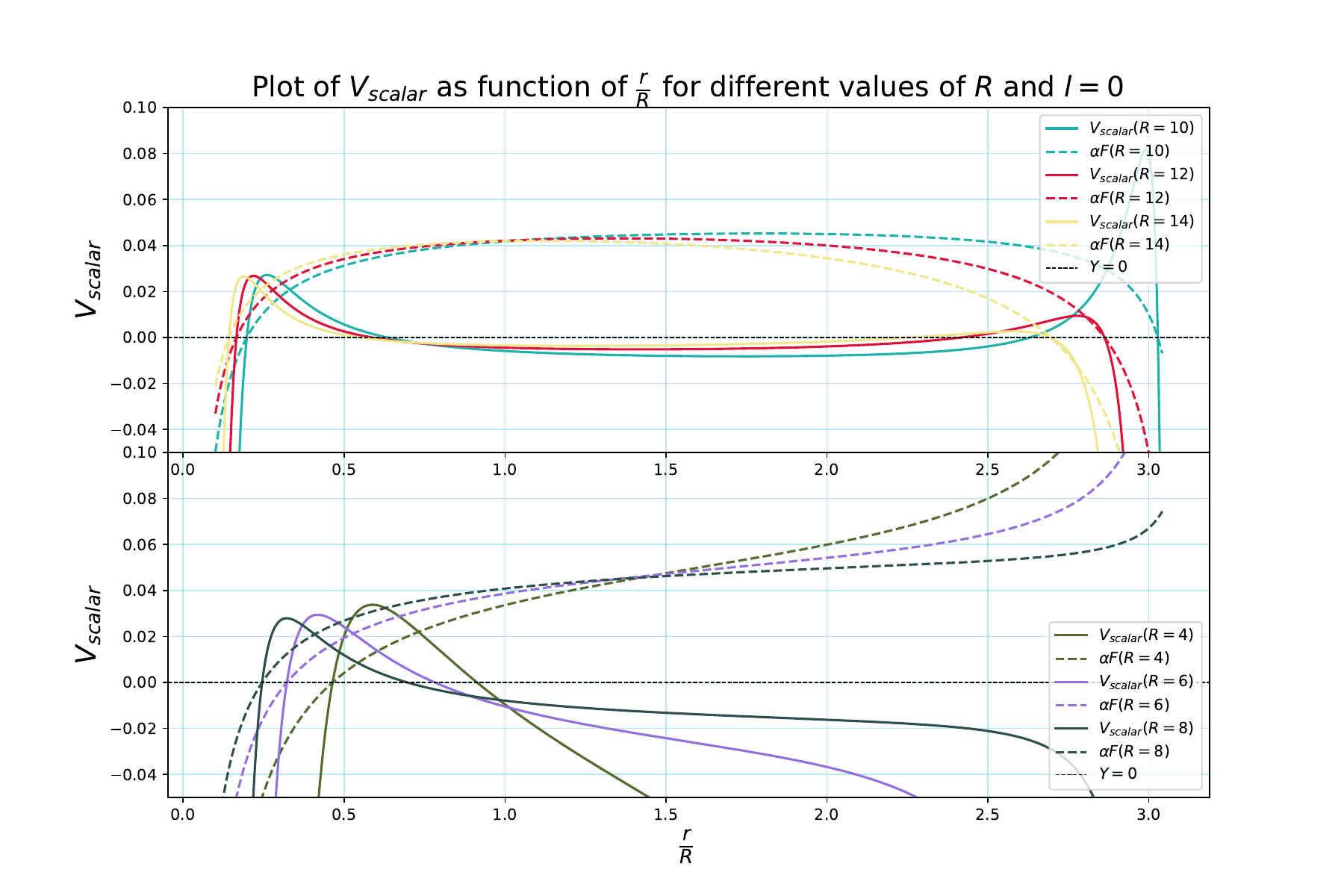}%
        \label{fig:VR2a}%
        }
    \subfloat[$V_{scalar}(r)$ for $l=1$]{%
        \includegraphics[scale=0.3]{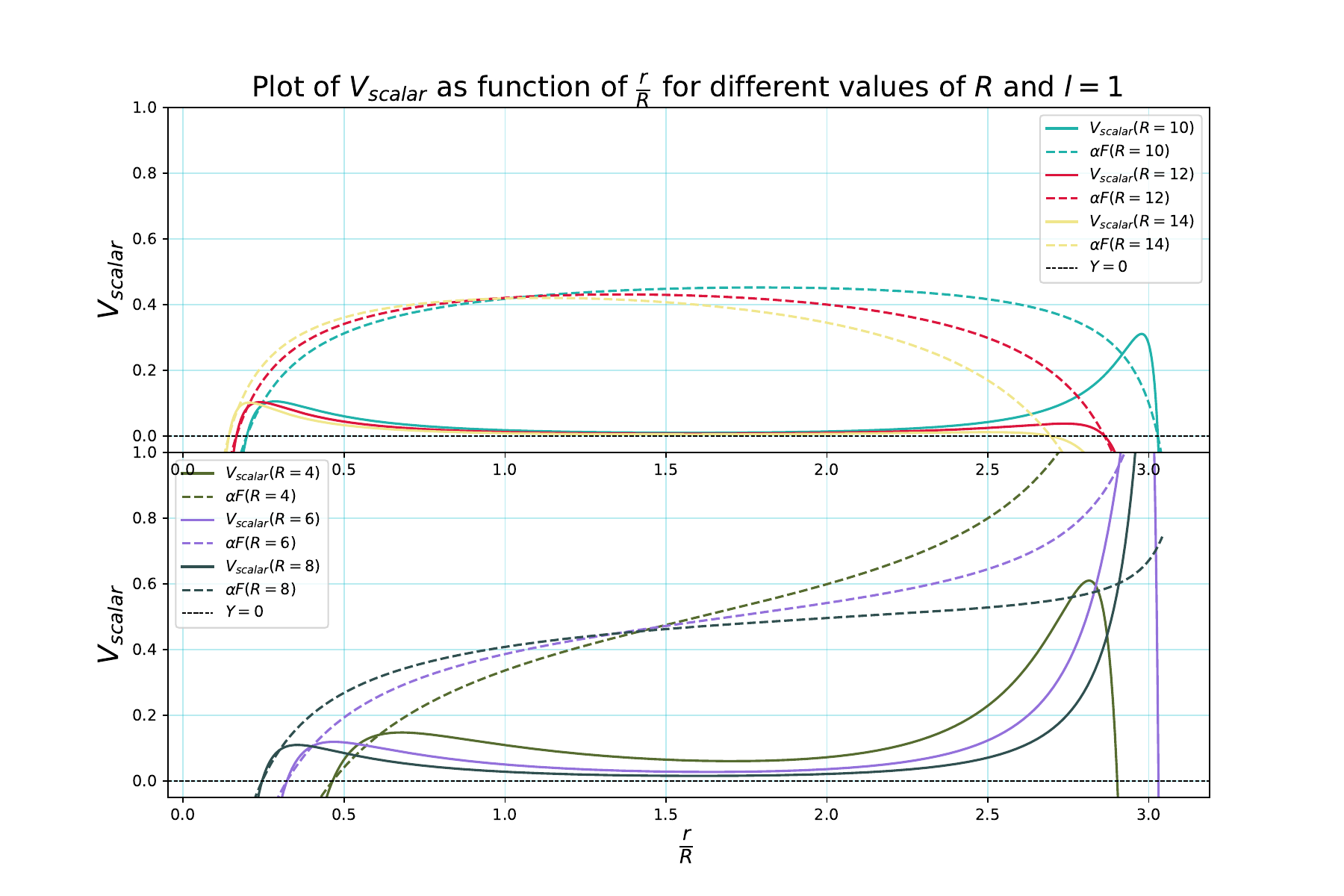}%
        \label{fig:VR2b}%
        }%
\caption{Plots of effective potential for scalar field as a function of $\frac rR$, for  $M=1$ and 
$\Lambda=0.001$ with various values of $R$ from $5.0$ to $14.0$. 
\label{fig:VR2}}
\end{center}
\end{figure}
\section{Scalar Waves in Schwarzschild-Whittaker \label{sec:sch-wth} }
\begin{figure}[t]
\begin{center}
\subfloat[$V_{scalar}(r)$ for $l=0$ ]{%
	\includegraphics[scale=0.3]{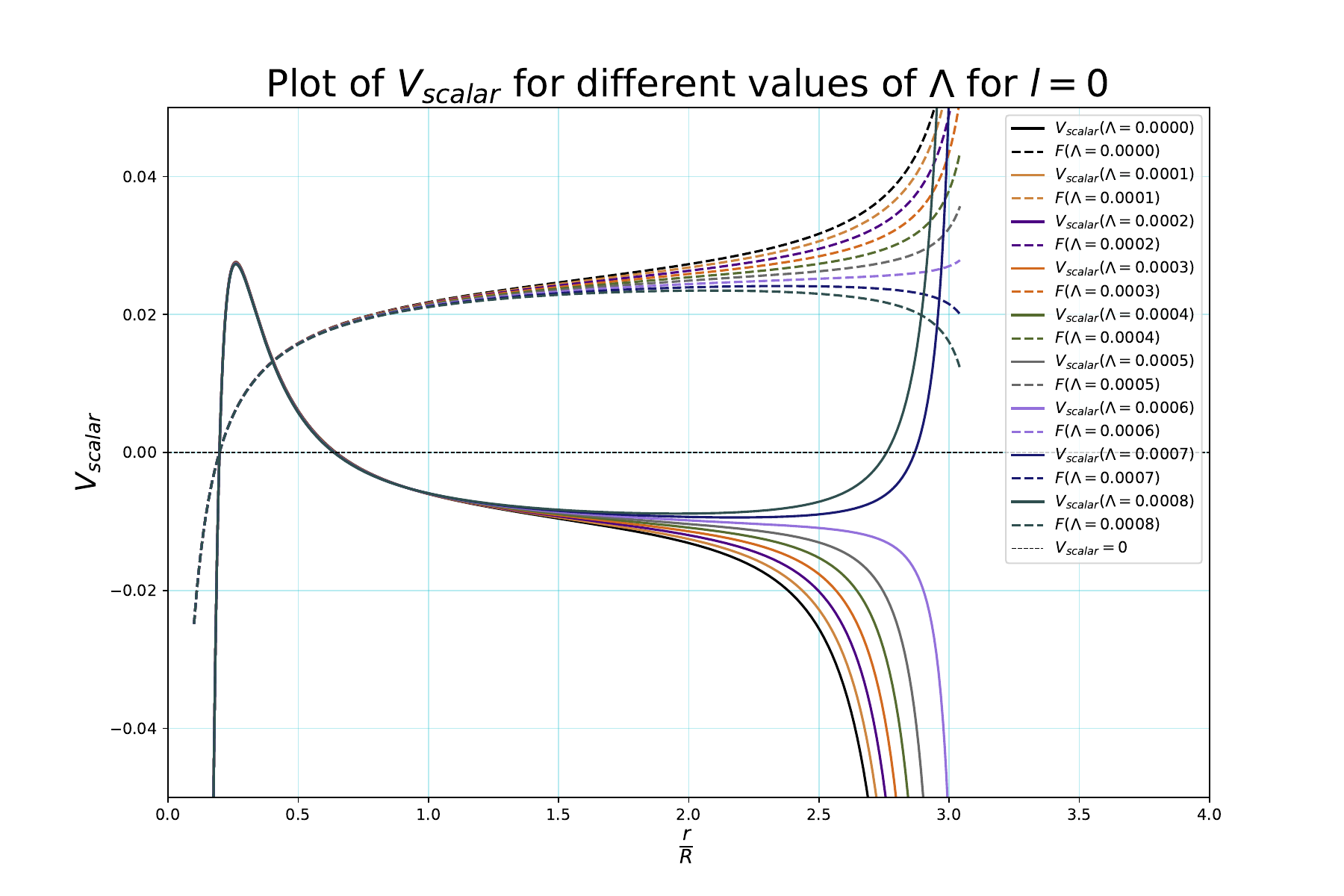}%
        \label{fig:VR3a}%
        }
    \subfloat[$V_{scalar}(r)$ for $l=1$]{%
        \includegraphics[scale=0.3]{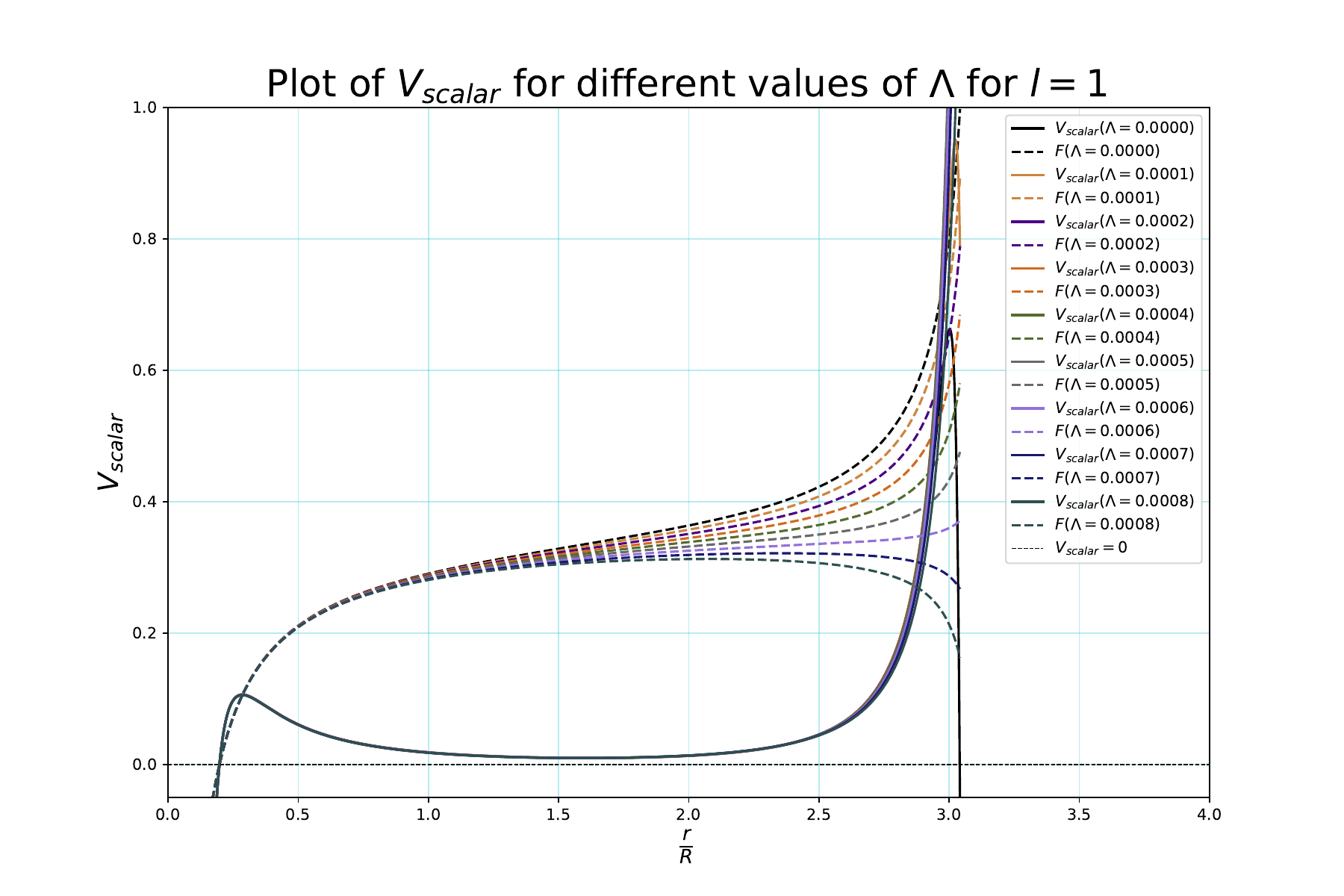}%
        \label{fig:VR3b}%
        }%
\caption{Plots of effective potential for scalar field as a function of $\frac rR$, for  $M=1$ and 
$R=10$. with various values of $ \Lambda$ from $0$ to $0.0008$. 
\label{fig:VR3}}
\end{center}
\end{figure}

In this section, we investigate a massless scalar field within the SW spacetime background. The evolution of the massless scalar field in a curved geometry is described by the Klein-Gordon equation, which can be written as follows:
\begin{equation}
\frac{1}{\sqrt{-g}}\frac{\partial}{\partial x^{a}}\left[\sqrt{-g}\:g^{ab}\frac{\partial\psi}{\partial x^{b}}\right] \, =\, 0\,.\label{eq:K-G}
\end{equation}
 Where $g$ is the determinant of the metric tensor $g_{ab}$, and for the spacetime metric given by eq.~(\ref{eq:ds2}), $g$ takes the expression $|g| = R^{4}\sin^{4}\left(\frac{r}{R}\right)\sin^{2}\theta$. The Klein-Gordon equation can be simplified to the following form:
\begin{equation}
\frac{1}{R^{2}\sin^{2}\left(\frac{r}{R}\right)\sin\theta}\frac{\partial}{\partial x^{a}}\left[R^{2}\sin^{2}\left(\frac{r}{R}\right)\sin\theta \, g^{ab}\frac{\partial\psi}{\partial x^{b}}\right]=0\,.
\end{equation}
 Following the standard procedure, we decompose the $\left(\theta,\phi\right)$ part of $\psi(t,r,\theta,\phi)$ in terms of spherical harmonics, and the temporal part in terms of the Fourier frequency $\omega$, resulting in the form:
\[
\psi\left(t, r,\theta,\phi\right) = e^{i\omega t}\,\psi_{r}\left(r\right)\,Y_{l}^{m}\left(\theta,\phi\right)\,,
\]
where $Y_{l}^{m}\left(\theta,\phi\right)$ are spherical harmonics. With this decomposition, we can separate the radial equation, which can be written as:

\begin{equation}
\frac{F(r)}{R^{2}\sin^{2}\left(\frac{r}{R}\right)}\frac{d}{dr}\left[R^{2}\sin^{2}\left(\frac{r}{R}\right)F\left(r\right)\frac{d\psi_r\left(r\right)}{dr}\right]+\left[\omega^{2}\,-\,\frac{l\left(l+1\right)F(r)}{R^{2}\sin^{2}\left(\frac{r}{R}\right)}\right]\psi_r\left(r\right)=0\,.
\end{equation}
We further simplify the equation by changing our analysis to a tortoise-like coordinate system. We define a new variable $y$ such that,
\begin{equation}
dy=F\left(r\right)^{-1}dr\,,
\end{equation}
and changing dependent variable from $\psi_r$ to $u(r)$, which is given by,
\begin{equation}
u\left(r\right)=R\sin\left(\frac{r}{R}\right)\psi_r\left(r\right)\,.
\end{equation}
With the above substitution, the radial equation takes the form of a time-independent Schr\"{o}dinger equation,
\begin{equation}
\frac{d^{2}u}{dy^{2}}\,+\,\left[\omega^{2}-V_{scalar}\left(r\right)\right]u=0\,,
\end{equation}
with the potential $V_{scalar}(r)$ given by,
\begin{equation}
V_{scalar}\left(r\right)=F(r)\left[\frac{1}{R\,\tan\left(\frac{r}{R}\right)}\frac{dF\left(r\right)}{dr}-\frac{1}{R^{2}}F\left(r\right)+\frac{l\left(l+1\right)}{R^{2}\sin^{2}\left(\frac{r}{R}\right)}\right]\,.
\end{equation}
Note that the potential, $V_{scalar}(r)$, vanishes on the event horizon, which is given by the condition $F\left(r_{h}\right)=0$. Plots of the scalar potential, $V_{scalar}(r)$, as functions of various parameters are shown in Figures \ref{fig:VR1}, \ref{fig:VR2}, and \ref{fig:VR3}. One important difference from the Schwarzschild case is that the potential takes negative values. The negative value of the potential on the boundary implies an exponential growth in the scalar field, leading to instability. However, at a finite radial point with negative potential, the field can have bound states, provided the fields have energies in a suitable range. If the energy level is kept suitably high, one can have the scattering of scalar field by the horizon. In addition, if the potential is negative beyond the horizon, this region is outside our domain.

 In Figure \ref{fig:VR1}, we display the potential for different values of $l$ ranging from $0$ to $6$ as a function of $\frac{r}{R}$, with $R=10$, $M=1$, and $\Lambda=0.001$ held constant. In this case, the potential is well-behaved, approaching zero on the boundary and beyond. For the lowest value of $l=0$, the minimum value inside the domain falls below zero, indicating the possibility of bound states between the blackhole and cosmological horizon. This could be attributed to a radial mode or spherical accretion of the field,  influenced by the presence of two horizons, suggesting that the free fall of matter may be affected by both horizons. 

Another interesting observation is the presence of two peaks, each closer to the blackhole and cosmological horizon. This configuration allows the field to become trapped between these peaks. However, the behaviour of the potential is strongly dependent on other parameter ranges due to the metric's dependence on multiple parameters.

Next, we investigate the influence of $R$ on the effective potential $V_{scalar}(r)$. The results are depicted in Figure \ref{fig:VR2} for two values of $l$ ($0$ and $1$). The metric coefficient $g_{00}$, denoted as $F(r)$, undergoes significant changes when altering the values of $R$ or accounting for cosmological influence. In these plots, $g_{00}=F(r)$ is also shown with a suitable scaling factor, $\alpha$, to bring the plots within the range. This is done to emphasize the regions where $F(r)>0$ and $V_{scalar}$. 

For lower values of $R=\{4, \, 6, \, \rm{and} \, 8\}$ (as shown in Figure \ref{fig:VR2a}), only the blackhole horizon exists, and there is no cosmological horizon. In all these cases, $V_{scalar}$ approaches negative values in the outer region, while $F(r) > 0$. These parameters indicate a potentially unstable spacetime and require more in-depth study.

Conversely, for larger values of $R=\{10, \,12, \, \rm{and} \, 14\}$ (Figure \ref{fig:VR2a}), the potential exhibits dual peaks. In these cases, both blackhole and cosmological horizons coexist, where the potential vanishes. Here, bound states are possible as the potential takes on negative values.

We also provide plots for $l=1$ in Figure \ref{fig:VR2b}. In Figure \ref{fig:VR3}, we present plots of $V_{scalar}$ for different values of $\Lambda$ while keeping other parameters fixed. For $l=0$ and $\Lambda=\{0.0007,\, 0.0008\}$, the potential blows up at the cosmological horizon. In these scenarios, the scalar field might be able to evolve and necessitates further detailed investigation. Figure \ref{fig:VR3a} illustrates that these two cases feature both horizons. 

However, for the remaining values of $\Lambda$, only the blackhole horizon is present, and the potential assumes negative values at the outer boundary, potentially indicating an unstable spacetime. In contrast, for $l=1$, the potential becomes zero at the cosmological horizon (Figure \ref{fig:VR3b}). It is evident that the behaviour of the massless scalar field in the SW solution is complex and warrants a more detailed analysis of stability, a subject we intend to address in the future.
\section{Conclusions \label{sec:con}}
In this article, we explore solutions to Einstein's equations with an event horizon surrounded by matter and asymptotically not flat. This is a generalization of the solution proposed by Vaidya~\cite{Vaidya1977} and falls under a special case of the Whittaker equation of state, which we refer to as the Schwarzschild-Whittaker solution. In this case, we have two independent parameters that induce cosmological effects: the size of the universe, denoted as $R$, and the cosmological constant $\Lambda$. We also extend the solution to a radiating Vaidya-like solution, incorporating a timelike perfect fluid with ingoing/outgoing radiation fields. We investigate various physical phenomena, such as the precession of periastron and the scattering of a massless scalar field, to understand the effects of cosmological parameters. Our preliminary studies suggest that the size of the universe and the cosmological parameters have roughly opposite contributions to the physics of blackholes in a cosmological background. Further detailed investigations are necessary for a proper understanding of cosmological influences on blackholes.
\bibliographystyle{unsrt}
\bibliography{references}
\end{document}